\documentclass{aa}

\usepackage[utf8]{inputenc}
\usepackage[T1]{fontenc}
\usepackage{graphicx}
\usepackage{amsmath}
\usepackage{ulem}
\usepackage{soul}
\usepackage{float}
\usepackage[colorlinks=true,
            linkcolor=blue,
            citecolor=blue,
            urlcolor=blue]{hyperref}
\usepackage{placeins}

\newcommand{\nustar}{\textit{NuSTAR}}

\newcommand{\ixpe}{\textit{IXPE}}

\begin{document}

\title{Spectro-polarimetric study of the black hole X-ray binary Swift J1727.8-1613 in the JED--SAD framework}

\titlerunning{Spectro-polarimetry of Swift J1727.8$-$1613 in the JED--SAD framework}
\authorrunning{Muhammad et al.}

\author{
R. Muhammad\inst{1}\thanks{E-mail: rafan.muhammad@univ-grenoble-alpes.fr}
\and P.-O. Petrucci\inst{1}
\and W. Zhang\inst{2}
\and J. Ferreira\inst{1}
\and R. Belmont\inst{3}
\and J. Malzac\inst{4}
\and G. Marcel\inst{5}
\and G. Matt\inst{6}
\and M. Parra \inst{7}
\and T. Bouchet\inst{8}
\and N. Barnier\inst{9}
\and M. Clavel\inst{1}
\and G. Henri\inst{1}
}

\institute{
$^{1}$ Université Grenoble Alpes, CNRS, IPAG, Grenoble, France \\
$^{2}$ National Astronomical Observatories, Chinese Academy of Sciences, A20 Datun Road, Beijing 100101, China \\
$^{3}$ CEA, IRFU, Département d'Astrophysique, Gif-sur-Yvette, France \\
$^{4}$ Université de Toulouse, CNRS, IRAP, UPS, Toulouse, France \\
$^{5}$ Department of Physics and Astronomy, FI-20014 University of Turku, Finland \\
$^{6}$ Dipartimento di Matematica e Fisica, Università Roma Tre, Rome, Italy`\\ 
$^{7}$ Department of Physics, Ehime University, 2-5, Bunkyocho, Matsuyama, Ehime 790-8577, Japan \\
$^{8}$ Julius-Maximilians-Universität Würzburg, Fakultät für Physik und Astronomie, Institut für Theoretische Physik und Astrophysik, Lehrstuhl für Astronomie, Emil-Fischer-Str. 31, 97074 Würzburg, Germany \\
$^{9}$ International Center for Hadron Astrophysics, Chiba University, Inage-ku, Chiba 263-8522, Japan
}

\abstract
{Swift J1727.8-1613 is an X-ray binary that was observed multiple times by \textit{IXPE}, \textit{NICER}, \textit{NuSTAR}, and \textit{INTEGRAL} during its 2023-2024 outburst. It was observed in a hard state at the beginning and the end of the outburst and in a soft state in between.}
{We aim to reproduce the spectro-polarimetric signatures observed during two hard states: one bright at the beginning of the outburst and the other dim at the end. The two show similar X-ray polarimetric signatures, but  their X-ray fluxes differ by two orders of magnitude.}
{We modelled these hard state spectra in the JED--SAD framework, which combines a jet-emitting disk (JED) in the inner region of the accretion flow and a standard accretion disk (SAD) in the outer region. The X-ray polarisation properties of the best-fit JED--SAD configurations were then computed with the MONK general relativistic radiative transfer code.}
{The spectral characteristics of both the bright and dim hard states are well reproduced within the JED--SAD framework. Within our modelling, however, the dominant radiative process differs between the two states: self-Comptonised bremsstrahlung in the bright hard state and self-Comptonised synchrotron in the dim hard state. In the bright hard state, the resulting polarisation fraction remains below the observed value of $\sim4\%$.  Moreover, due to the JED's large vertical optical depth ($\sim$10), the polarisation angle is aligned with the accretion disk surface, i.e. there is a $90^\circ$ difference with the observations. In contrast, the dim hard state's simulated polarisation fraction is about 3\% and the polarisation angle is perpendicular to the disk, both of which are broadly consistent with the \ixpe measurements. Our simulations of the dim hard state are largely insensitive to the magnetic field configuration due to the strong depolarisation of the synchrotron seed photons by Faraday rotation.} 
{While the spectro-polarimetric properties of the dim hard state are well reproduced by our model, the bright hard state polarisation properties do not agree with the observations. We discuss different processes that could be responsible for the origin of the observed polarisation in the bright hard state: the presence of a non-thermal particle distribution, which would be  consistent with the presence of a hard tail observed above 50 keV; and the electron scattering in an accretion-disk wind, which could modify both the polarisation fraction and angle. We discuss the limitations of the current JED--SAD framework and outline possible improvements to better reproduce the \textit{IXPE} data.}

\keywords{X-rays: binaries -- stars: black holes -- accretion, accretion disks -- X ray polarisation}

\maketitle
\nolinenumbers

\section{Introduction} \vspace{5pt} \label{introduction}

Swift J1727.8-1613 is a transient X-ray binary system consisting of a stellar-mass black hole and a companion star. The first distance estimates put it at approximately 2.7 $\pm$ 0.3 kpc, while the newest distance estimations vary from 1.5--4.3 kpc \citep{hughes2025, mata-sanchez2024, burridge2025}. This source exemplifies the class of black hole X-ray binaries (BHXBs), which are known for extreme variability in X-ray luminosity and significant spectral state transitions. First identified during an intense outburst in August 2023, Swift J1727.8-1613 reached an X-ray luminosity of approximately 7 crab in the 2--20 keV energy band, marking it as an exceptionally bright BHXB system \citep{podgorny2024, veledina2023}.

Similar to previously studied systems such as GX 339-4, Swift J1727.8-1613 exhibits the canonical transitions between the hard and soft spectral states characteristic of stellar black hole accreting systems. Using the Imaging X-ray Polarimetry Explorer (\textit{IXPE}) telescope, eight observations were made during the Swift J1727.8-1613 outburst. These eight \textit{IXPE} epochs cover the Swift J1727.8-1613 hardness-intensity diagram, with five observations in the hard-to-soft transition, two in the soft state, and one in the dim hard state (after the soft-to-hard transition at the end of the outburst). Although during the first \textit{IXPE} observation Swift J1727.8-1613  was already in the early hard-to-soft transition, we refer to this epoch as the ‘bright hard state’ throughout this work.

The measured polarisation fraction in the hard-to-soft states remains roughly constant across all five \textit{IXPE} epochs. The observations show a polarisation fraction of about 3\% at 2 keV and 5\% at 8 keV, with a linear increase across the 2--8 keV energy band, and a polarisation angle that remains approximately constant, centred around $0^\circ$ \citep{ingram2024, veledina2023}. The polarisation angle is consistent with the jet position angle resolved by radio observations \citep{wood2024}, similarly to other BHXB hard states observed by \textit{IXPE}, like Cygnus X-1 \citep{krawczynski2022} and  GX 339-4 \citep{mastroserio2025}.
However, during the transition to the soft state, the polarisation fraction drops, with an upper limit of 1.2\% \citep{svoboda2024}. The source returns to the hard state at the end of the outburst, when the \textit{IXPE} measurements showed a polarisation fraction and angle signatures similar to those observed in the bright hard state, despite a difference of two orders of magnitude in luminosity \citep{podgorny2024}. \\

We aimed to reproduce the spectro-polarimetric properties of the Swift J1727.8-1613 hard states in the JED--SAD framework \citep{ferreira2006}.  This framework assumes the presence of a large-scale magnetic field that threads the accretion flow, dividing it radially into a highly magnetised inner region and producing a jet, the so-called jet-emitting disk (JED) framework. This\ JED is surrounded by a weakly magnetised outer region, the so-called standard accretion disk (SAD), which is located beyond a transition radius ($R_{j}$; also called $R_{tr}$ in previous papers). Such a radial distribution of the magnetisation is now observed in global magnetohydrodynamic (MHD) simulations (see e.g. \citealt{scepi2024a, scepi2024b, Liska2022, zimniak2026}) and results from the advection and diffusion of the magnetic field inside the accretion flow. The JED--SAD model has been successfully used to reproduce the X-ray spectra of several compact objects \citep{petrucci2010,marcel_unified_2018-1,marcel_unified_2018,marcel_unified_2019,marino2021,barnier2022,marcel2020,marcel2022}. 

We investigated the polarisation properties of the JED--SAD model when applied to the hard states of  Swift J1727.8-1613. This was done by using  the radiative transfer simulation code MONK \citep{zhang2019}. This C++-based Monte Carlo code   allows the user to compute the spectral and polarisation signatures of any geometry. Combining the JED--SAD model with MONK provides a preliminary framework for exploring the spectral and polarimetric properties of transient X-ray binaries such as Swift J1727.8-1613.\\

This paper is organised as follows. We present the data of Swift J1727.8-1613 used in our JED--SAD fit in Sect. \ref{obsSect}. The main characteristics of the JED--SAD framework are recalled in Sect. \ref{jedsadsect}, and the MONK code properties are presented in Sect. \ref{monksect}. The spectral and polarisation results are detailed in Sect. \ref{ResultSect} and discussed in Sects. \ref{DiscSect} and \ref{Sec:Energy_budget_caveats}. We also compare the spectra of the DYPLO and MONK codes in Appendix~\ref{Appendix}.

\begin{table*}
\caption{Observation IDs used for the spectral fitting.}
\label{tab:obsids}
\centering
\begin{tabular}{lcccccc}
\hline\hline
Observation date & Spectral state & \textit{IXPE} & \textit{NICER} & \textit{NuSTAR} & \textit{INTEGRAL} & \textit{Swift} XRT/BAT \\
\hline
Sept. 7, 2023 & Bright hard & 02250901 & 6750010501 & 80902333006 & 2683 & -- \\
Apr. 3, 2024 & Dim hard    & 03005801 & 7708010109 & -- & -- & 00089766012 \\
\hline
\end{tabular}
\end{table*}

\section{Observation and data reduction}
\label{obsSect}
Swift J1727.8-1613 underwent a major outburst between August 2023 and June 2024, reaching a peak X-ray flux of $\sim$7 crab and evolving through the canonical spectral states of black hole transients. During the state transitions, strong low-frequency quasi-periodic oscillations and a transient radio jet were detected \citep{mereminskiy2024, wood2024, hxmt2025swiftj1727}. In this work, we focused on two hard-state observations obtained with \textit{IXPE}. The first was obtained on September 7, 2023, during the bright hard state. The second, obtained on April 4, 2024, after the return to the hard state, corresponds to the dim hard state. The X-ray flux of the latter is approximately two orders of magnitude lower than that of the former.

Each dataset combines simultaneous observations from multiple observatories (Table~\ref{tab:obsids}). The bright hard state dataset includes \textit{NICER}, \textit{NuSTAR}, \textit{INTEGRAL}, and \textit{IXPE} observations, whereas the dim hard state dataset combines \textit{NICER}, \textit{Swift}/XRT, \textit{Swift}/BAT, and \textit{IXPE}. All data were reduced using HEASOFT v6.33.2.

\subsection{\textit{IXPE} observations}

The bright hard state (ObsID: 02250901) of Swift J1727.8-1613 was observed by \textit{IXPE} \citep{weisskopf2022ixpe} from September 7 to 8, 2023, with a live time of $\sim$19 ks \citep{dovciak2023a,dovciak2023b,veledina2023}. The dim hard state (ObsID: 03005801) was observed between April 3 and 8, 2024, with a total live time of 202 ks \citep{podgorny2024}. The Level-2 data were retrieved from the HEASARC archive and reduced using IXPEOBSSIM \citep{baldini2022} and HEASOFT/XSPEC \citep{arnaud1996}, following \citet{veledina2023} and \citet{podgorny2024} for the bright and dim hard states, respectively. Our measurements are consistent with these studies: both observations show a polarisation fraction increasing from $\sim$3\% at 2 keV to $\sim$5\% at 8 keV and a polarisation angle aligned with the radio jet axis \citep{wood2024}. Similar signatures are observed in other hard-state BHXBs, such as Cyg X-1 \citep{kravtsov2025} and GX 339--4 \citep{mastroserio2025}; see also \citet{dovciak2024} and \citet{marcel2026}.

\subsection{\textit{NICER} observations}
The Neutron Star Interior Composition Explorer (\textit{NICER}; \citealt{Gendreau2016_NICER}) observed Swift J1727.8-1613 from August 25 to October 9, 2023. The data were reduced using NICERDAS\footnote{\href{https://heasarc.gsfc.nasa.gov/docs/nicer/nicer_analysis.html}{https://heasarc.gsfc.nasa.gov/docs/nicer/nicer\_analysis.html}} version 12 and \textit{NICER} CALDB xti20240206, following the standard analysis threads. The geomagnetic data required by the \texttt{scorpeon} background model were obtained using the \texttt{nigeodown} task. Due to the exceptional brightness of Swift J1727.8-1613, we adopted a fixed undershoot threshold of 800 counts\,s$^{-1}$. High-background intervals were identified by comparing the 2--8\,keV count rate with the overshoot rate and subsequently removed. Remaining non-X-ray flares were filtered using a combination of topological and variance-based peak detection algorithms. The background was estimated with the \texttt{scorpeon} model, and the spectra were grouped following the optimal binning prescription of \citet{Kaastra2016_binning_opt}. Due to enhanced background contributions and residual calibration features below 0.6\,keV, the spectral analysis was restricted to the 0.6--10\,keV energy range.

\subsection{NuSTAR observations}

The Nuclear Spectroscopic Telescope Array (\nustar{}; \citealt{Harrison2013_NuSTAR}) observed Swift J1727.8-1613 on September 7, 2023 (ObsID: 80902333006). The data were reduced using the standard \texttt{NuSTARDAS} tasks\footnote{\href{https://heasarc.gsfc.nasa.gov/docs/nustar/analysis/}{https://heasarc.gsfc.nasa.gov/docs/nustar/analysis/}} and \nustar{} CALDB v20230613. Following the recommendations for bright sources\footnote{\href{https://heasarc.gsfc.nasa.gov/docs/nustar/nustar_faq.html\#bright}{NuSTAR analysis recommendations for bright sources}}, the \texttt{nupipeline} processing was performed using the ``\texttt{(STATUS==b0000xxx00xxxx000)\&\&(SHIELD==0)}'' screening criterion. Source and background extraction regions were defined following the signal-to-noise optimisation method of \citet{Piconcelli2004_SNR_opti}. Due to the exceptional brightness of Swift J1727.8-1613, the optimal extraction radius was found to be 120\arcsec. Spectral products were extracted independently for the two focal plane modules FPMA and FPMB using \texttt{nuproducts} and grouped following the optimal binning prescription of \citet{Kaastra2016_binning_opt}. Since no significant discrepancies were found between the two focal plane modules, no multi-layer insulation (MLI) correction was applied and a free cross-normalisation constant was included in the spectral fits. The spectral analysis was performed over the 4--78\,keV energy range.

\subsection{\textit{INTEGRAL} observations}

The International Gamma-Ray Astrophysics Laboratory (\textit{INTEGRAL}) satellite observed the source in the bright hard state. We relied on the Imager on-Board the \textit{INTEGRAL} Satellite \citep[IBIS;][]{Ubertini_2003}, in particular its upper detection plane, the IBIS Soft Gamma Ray Imager \citep[ISGRI;][]{Lebrun_2003}. The spectrum was extracted with the Off-line Scientific Analysis (OSA) software  v11.2, following standard reduction processes\footnote{\url{https://www.isdc.unige.ch/integral/download/osa/doc/11.2/osa_um_ibis/man_html.html}}, and as described in \cite{bouchet2024}. We used the combined observations from the satellite revolution 2683 (MJD\,60194.2-60195.2), resulting in 64\,ks of exposure, and restrained the energy range to $60-150$\,keV.

\subsection{Swift observations}

We used the public \textit{Swift} X-Ray Telescope (XRT) observation 00089766012, obtained on April 3, 2024, in windowed timing mode. The source spectrum was extracted together with the corresponding background, response matrix, and ancillary response files following standard \textit{Swift}/XRT analysis procedures \citep{burrows2005xrt}. The resulting spectrum has a net exposure of 1.65 ks and was fitted over the 0.6--10 keV energy range. We also included the corresponding \textit{Swift} Burst Alert Telescope (BAT) spectrum, with a net exposure of 71.2 ks, to extend the spectral coverage towards hard X-ray energies \citep{barthelmy2005bat}. The XRT and BAT spectra were fitted simultaneously within \textsc{xspec} \citep{arnaud1996}.

\section{The framework}
\subsection{The JED--SAD model}
\label{jedsadsect}
The JED--SAD model, introduced in several studies \citep{ferreira2006, marcel_unified_2018-1, marcel_unified_2018, marcel_unified_2019}, provides a semi-analytical description of accretion flows in which a large-scale magnetic field plays a central role. This field drives jet production and exerts a torque on the accretion disk, extracting angular momentum vertically. As a result, the accretion flow can accelerate to transonic or even supersonic speeds while remaining optically thin and hot, even at luminosities exceeding 10\% of the Eddington limit.  

In the JED--SAD framework, it is assumed that the radial distribution of the magnetic field produces a two-phase radial accretion structure: an inner, strongly magnetised JED, and an outer, weakly magnetised SAD (\citealt{shakurasunyaev73}), beyond a transition radius ($R_{j}$). The JED radial structure is obtained within a parameterised MHD framework, whose parameter space is motivated by the self-similar accretion-ejection solutions developed by Ferreira  (\citeyear{ferreira1997}). Such radial distributions are in agreement with those seen in several general relativistic MHD simulations (e.g. \citealt{Liska2022, scepi2024a,scepi2024b}). \\ 

The JED--SAD thermal structure and spectral emission are computed with a python code named DYPLO \citep{marcel_unified_2018}. This code solves the thermal equilibrium at each radius, balancing the accretion power release with radiative cooling, outflows and advection. We also assumed a two-temperature plasma with an identical amount of accretion power heating the protons and the electrons (see \citealt{marcel_unified_2018} for more details).
To compute the corresponding emitted spectrum, \citet{marcel_unified_2018, marcel_unified_2018-1} used the \texttt{belm} radiative code \citep{belmont_simulating_2008}, which accounts for synchrotron self-absorption, bremsstrahlung, and Compton scattering. Radiative processes in \texttt{belm} are computed assuming a spherical geometry. To be consistent with this assumption, DYPLO divides the JED into rings, each populated with spherical emitting regions. Radiative emission can be computed in both the optically thin and optically thick regimes; however, the transition between these two states is not trivial. Therefore, DYPLO uses a bridge formula to account for the intermediate regime (see e.g. \citealt{marcel_unified_2018}). The emitting spectrum is then the sum of all the spectra computed at different radii in the JED--SAD configuration.  

The radial dynamical structure of the JED--SAD solution is determined by a set of parameters (see \citealt{marcel_unified_2018}). They include dynamical parameters, namely the accretion Mach number ($m_s$), the SAD viscosity parameter ($\alpha_\nu$), the ejection efficiency ($p$), the disk magnetisation ($\mu$), and the jet power fraction ($b$). They also include astrophysical parameters, namely the black hole mass ($M_{\rm BH}$), the accretion rate at the innermost stable circular orbit (ISCO; $\dot M_{\rm in}$, denoted simply $\dot M$ in the following), the radius of the innermost stable circular orbit ($R_{\rm ISCO}$), the JED--SAD transition radius ($R_J$), the outer SAD radius ($R_{\rm out}$), and the geometrical parameter ($\omega$), which controls the fraction of SAD photons Comptonised within the JED. We stress that DYPLO does not include relativistic effects: therefore, varying $R_{\rm ISCO}$ should not be interpreted as a direct test of the black hole spin. All the dynamical parameters described above are consistent with the values derived in full MHD calculations of the accretion-ejection dynamics (see e.g. \citealt{marcel_unified_2018}). Thus, although the dynamics are not computed when calculating the disk thermal states and spectra, the latter are confidently consistent with the former (see e.g. \citealt{marcel2022}). The only unknowns when fitting the X-ray data are the following three parameters : $R_J$, $\dot{M}$, and $m_s$. Note that in this framework the accretion rate is expressed in units of $L_{Edd} / c^2$, which differs from the conventional Eddington-scaled accretion rate by the radiative efficiency $\eta$. The black hole mass is fixed to $10\,M_{\odot}$, in agreement with independent measurements, although it remains poorly constrained for Swift J1727.8$-$1613 \citep{matasanchez2025}. This parameter has only a weak influence on the JED--SAD spectral results. $R_{\rm ISCO}$ is fixed between 1 and 6 $R_g$, while $R_{\rm out}$, $\alpha_{\nu}$, $p$, $\mu$, $b$ are set to fiducial values discussed in \cite{marcel_unified_2018}. Note that from Monte Carlo estimates, $\omega$ is generally expected to be small for an outer geometrically thin SAD and is fixed to 0.01 in the following. The spectral flux is then normalised by the source distance, which is generally constrained by independent measurements.

The JED--SAD model has been shown to reproduce the X-ray spectra of several accreting stellar-mass compact objects, like GX 339-4 \citep{marcel_unified_2019, barnier2022} and MAXI J1820+070  \citep{marino2021}, as well as active galactic nuclei such as HE 1143-1810 \citep{ursini2020}. Beyond the best-fit parameters, the model also provides the radial profiles of several key physical properties of the accretion flow, like the density, optical depth, temperature, or magnetic field strength.

\subsection{MONK}
\label{monksect}

MONK is a general relativistic Monte Carlo radiative transfer code designed to model the emission, absorption, and propagation of photons in curved space-time \citep{zhang2019}. 
It accounts for the main radiative processes expected in accretion flows, including thermal bremsstrahlung and synchrotron radiation, together with their respective self-absorptions. Compton scattering is also implemented, enabling a complete description of photon interactions in hot plasmas.  
\vspace{-0.05cm}
The code follows the Monte Carlo `super-photon' approach, in which each photon packet is assigned a weight corresponding to the number of physical photons it represents. Super-photons are generated in the local fluid frame according to the relevant emission coefficients, and their trajectories are then propagated along null geodesics by solving the geodesic equation in curved space-time.  

A key feature of MONK is its consistent treatment of polarisation. Each super-photon carries a polarisation vector that is evolved together with its wave vector, as well as the linear and circular polarisation degrees. These are used to derive the Stokes parameters (I, Q, U, V) in the rest frame of the fluid, which describes the polarisation state and is transported according to the polarised radiative transfer equation: 
\vspace{-0.05cm}
\begin{equation}
\frac{d\mathbf{I}}{d\lambda} = \mathbf{J} - \mathbf{M}\mathbf{I},
\end{equation}
where $\lambda$ is the affine parameter along the photon geodesic, $\mathbf{I} = (I, Q, U, V)$, $\mathbf{J}$ represents the emissivities, and $\mathbf{M}$ is the Mueller matrix \citep{dexter_public_2016} that encodes absorption, Faraday rotation, and Faraday conversion in magnetised plasma. In particular, synchrotron emission in ordered fields leads to intrinsically polarised radiation, whereas turbulent fields imply non polarised emission. Faraday effects are included through rotation and conversion coefficients that alter the polarisation angle and mix linear and circular components.  

By evolving the Stokes parameters along photon trajectories, MONK captures the combined effects of curved space-time, magnetic fields, and plasma opacities on both the intensity and polarisation of the emerging radiation. The full set of equations, together with the numerical methods used to solve them, are presented in detail in \cite{zhang2019} and Zhang et al. (in prep.). The latter also presents the application of the JED--SAD and MONK code to spectral and polarisation measurements of the high-mass BHXB  Cygnus X-1.

\section{Results}
\label{ResultSect}

\subsection{Bright hard state}
\label{Sec:BrightHardState}
\subsubsection{JED spectral fit}

\begin{figure}
\includegraphics[width=\columnwidth]{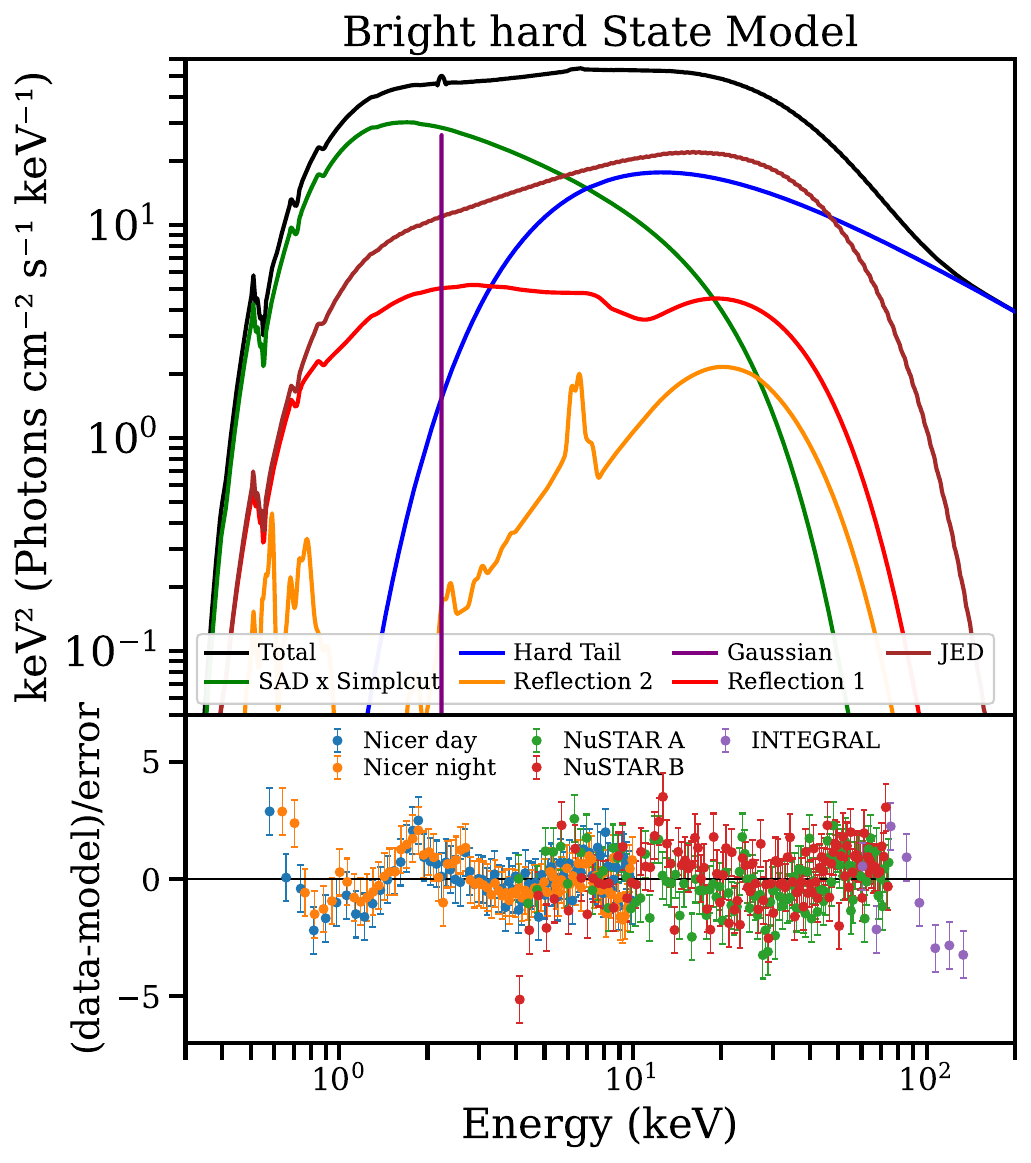}
 \caption{SED of the bright hard state best fit obtained using model \ref{model3}. The best-fit parameters are reported in Table \ref{table:bestfitmodelparameter} (left column) and the residuals at the bottom of Fig. \ref{fig:bright_preliminary_model}. The SAD is in green, JED in brown, Reflection 1 in red, Reflection 2 in orange, Gaussian in purple, and hard tail in blue. \label{fig:bright_model}}
\end{figure}

The bright hard state was observed by \textit{IXPE}, \textit{NICER}, \textit{NuSTAR}, and \textit{INTEGRAL}. We started the fitting with a model that includes the JED and SAD emission as well as a relativistically blurred reflection. The JED and SAD components are implemented using {\sc{xspec}}  tables \citep{barnier2022}. The reflection component is modelled with RelxillCp, using as inner radius the transition radius $R_{J}$ between the JED and the SAD. Since the JED component does not include a photon index parameter, we estimated the photon index by fitting the best-fit JED spectrum with an nthcomp component and used the resulting nthcomp parameters (photon index and high energy temperature) in RelxillCp.  We also accounted for the interstellar absorption  through the multiplicative model  {\sc{tbabs}} of {\sc{xspec}}. During the fitting procedure, we tested different ISCO radius values ($R_{\rm ISCO} = 2, 4,$ and $6$ gravitational radii, $R_g$). Both states give a better fit with $R_{\rm ISCO}$ = 4 $R_g$, and we therefore fixed $R_{\rm ISCO}$ to this value. Moreover, the black hole mass is also fixed to 10 solar masses and the JED magnetisation ($\mu$) to 0.5. We also included a narrow Gaussian component at 2.2 keV to account for instrumental residuals in the \textit{NICER} spectrum. Hereafter, the residuals of all spectral fits are shown in Appendix \ref{Appendix_bestfit}. In {\sc{xspec}} notation, the model (model~\ref{model1}) writes 
\vspace{-0.1 cm}
\begin{equation}
    \mathrm{Tbabs * (SAD + RelxillCp + JED)\label{model1}}
.\end{equation}

The spectral best fit obtained with this model gives a \(\chi^2\) = 20618 with 805 degrees of freedom, which is a poor fit. The fit has a clear discrepancy at high energy ($>80$ keV). 
This excess of emission with respect to the thermal Comptonisation component  (here modelled with the JED but usually modelled by a simple cutoff power law) has been observed in previous spectral studies (e.g. \citealt{yang2024,bouchet2024,mereminskiy2024}). This component is generally called the `hard tail'. Its origin is not clearly understood, but it could be the signature of non-thermal particles in the corona or the jet \citep{romero2014, joinet2007, delsanto2013, motta2021}. We thus phenomenologically added a power law to the model to account for this hard tail and improve the fit. We added a low energy cutoff ({\sc{expabs}} model in {\sc{xspec}}) to this power law to avoid an excess of emission at low energy. The new model (model~\ref{model2}) then becomes \begin{equation}
    \mathrm{Tbabs * (SAD + RelxillCp + JED + expabs*powerlaw)\label{model2}}
.\end{equation}

The fit improvement is highly significant with  \(\Delta \chi^2\) = 10323 for 3 additional parameters. As expected, the addition of the power law helps fit the data above $\sim$100 keV, in the \textit{INTEGRAL} energy range.\\

There are still large residuals present at low energies, below 5 keV, however. In this range the SAD and the JED dominate. The reflection is also present, with high ionisation parameter ($\log \xi$=3.3), but at a low level since the iron line and the reflection hump are not strong in the data. The fit is biased by the low-energy residuals, leading {\sc{xspec}} to increase the contribution of the SAD component through a reduction of the transition radius $R_J$. But the decrease in $R_J$ also modifies the JED spectrum, which becomes fainter as the decrease in $R_J$ leads to a reduction in the size of the JED, but also softer (Compton cooling by the SAD increases). A soft component is sometimes added to improve the fit at low energy of hard state spectra (e.g. \citealt{yamada2013,kolehmainen2014,basak2017,zdziarski2021,dzielak2021,zdziarski2025}). Its origin is not well known but it could be the counterpart of the soft X-ray excess known to exist in most radio-quiet active galactic nuclei (e.g. \citealt{arnaud1985,walterfink1993,page2004,gierlinski2004,crummy2006,bianchi2009,petrucci2018}) and whose origin could be thermal Comptonisation in a warm scattering layer above the optically thick accretion disk (e.g. \citealt{petrucci2020,ballantyne2020,xiang2022}). We therefore modified the model by convolving the SAD component with the {\sc{simplcut}}\footnote{https://jfsteiner.com/?p=71} component. We note that the best fit favours a low \textsc{simplcut} reflection fraction, which we fixed to zero. This is at odds with the $R_f\sim1$ expected for a slab-like warm corona \citep{steiner2017}. Fixing $R_f=1$ worsens the fit while leaving the other best-fit parameters broadly unchanged, questioning our physical interpretation of this additional component. To avoid edge effects inherent to the convolution process, the {\sc xspec} energy grid was extended from 0.01 to 1000 keV using logarithmically spaced energy bins. The current model (model~\ref{model3_prev}) is \begin{eqnarray}
    \mathrm{Tbabs * (simplcut * SAD + RelxillCp 1 + JED }&  \nonumber\\ \mathrm{+ expabs * powerlaw) \label{model3_prev}}
.\end{eqnarray}

Adding this component results in an improved fit, with $\chi^2 = 1182$ for 800 degrees of freedom, with a fit improvement of $\Delta\chi^2$ = 9113 for two additional degrees of freedom. However, significant residuals remain in the iron-line energy range around 6--7 keV. This indicates that the current model does not fully account for the observed reflection features. Thus, we included an additional reflection component to account for the reprocessing of the total emission in the outer regions of the SAD, in a configuration similar to the two-reflection scenario proposed for MAXI J1820+070 \citep{marino2021}. This second reflection component is hereafter referred to as RelxillCp2. The model (model~\ref{model3}) then becomes 
\begin{eqnarray}
    \mathrm{Tbabs * (simplcut * SAD + RelxillCp 1 + RelxillCp2} &  \nonumber\\  
     \mathrm{+ JED + expabs * powerlaw) \label{model3}}
.\end{eqnarray}

Using this model, our final best fit has $\chi^2$ = 951 for 798 degrees of freedom, with a fit improvement of $\Delta\chi^2$ = 231. The best-fit residuals and the best-fit spectral energy distribution (SED), together with the different spectral components, are shown in Fig. \ref{fig:bright_model}. The best-fit parameters are reported in the first column of Table \ref{table:bestfitmodelparameter}. We find a transition radius $R_J = 12.7 \pm 0.3\ R_g$, an accretion rate $\dot{M}=1.6 \pm 0.1$ $L_{Edd}/c^{2}$, a sonic Mach number $m_s=0.7 \pm 0.1$, and a normalisation of $16.8 \pm 0.5$, which corresponds to a distance of $2.4 \pm 0.1$~kpc.\ This is consistent with the most recent distance range of $1.5$--$4.3$~kpc adopted by \citet{hughes2025} based on a revised estimate by \citet{burridge2025} (from private communication).

\begin{table}
\caption{Best-fit parameters for the bright and dim hard states of Swift J1727.8$-$1613.}
\label{table:bestfitmodelparameter}
\centering
\setlength{\tabcolsep}{1pt}
\begin{tabular}{lcc}
\hline\hline
Parameter & Bright hard state & Dim hard state \\
\hline
$N_{\rm h}$ ($10^{22}\ \mathrm{cm}^{-2}$) & $0.37\pm0.1$ & $0.34\pm0.1$ \\
$R_{\rm J}$ ($R_{\rm g}$) & $12.7\pm0.3$ & $16.1\pm0.2$ \\
$\dot{M}$ ($L_{\rm Edd}/c^2$) & $1.6\pm0.1$ & $0.13\pm0.01$ \\
$m_{\rm s}$ & $0.7\pm0.1$ & $0.6\pm0.1$ \\
Normalisation & $16.8\pm0.5$ & 16.8 (fixed) \\
Reflection 1: $\log\xi$ & $3.3\pm0.1$ & $2.8\pm0.4$ \\
Reflection 2: $\log\xi$ & $0.1\pm0.1$ & -- \\
Hard-tail: $\Gamma$ & $2.8\pm0.1$ & -- \\
Hard-tail: cut off (keV) & $10.3\pm0.3$ & -- \\
{\sc simplcut} $\Gamma$ & $2.5\pm0.1$ & -- \\
{\sc simplcut} Scattering Frac & $1.0\pm0.1$ & -- \\
{\sc simplcut} $kT_{\rm e}$ (keV) & $6.0\pm0.6$ & -- \\
\hline
$\chi^2$/d.o.f. & 951/798 & 193/193 \\
\hline
\end{tabular}
\tablefoot{Fixed parameters are $b=0.3$,
$\alpha_{\rm v}=0.1$, $\mu=0.5$, $R_{\rm ISCO}=4$,
$M_{\rm BH}=10\,M_\odot$, $\omega=0.01$, and $p=0.01$.}
\end{table}

\subsubsection{Radiative properties of the jet-emitting disk solution} \label{sec412}
\begin{figure}
\includegraphics[width=\columnwidth]{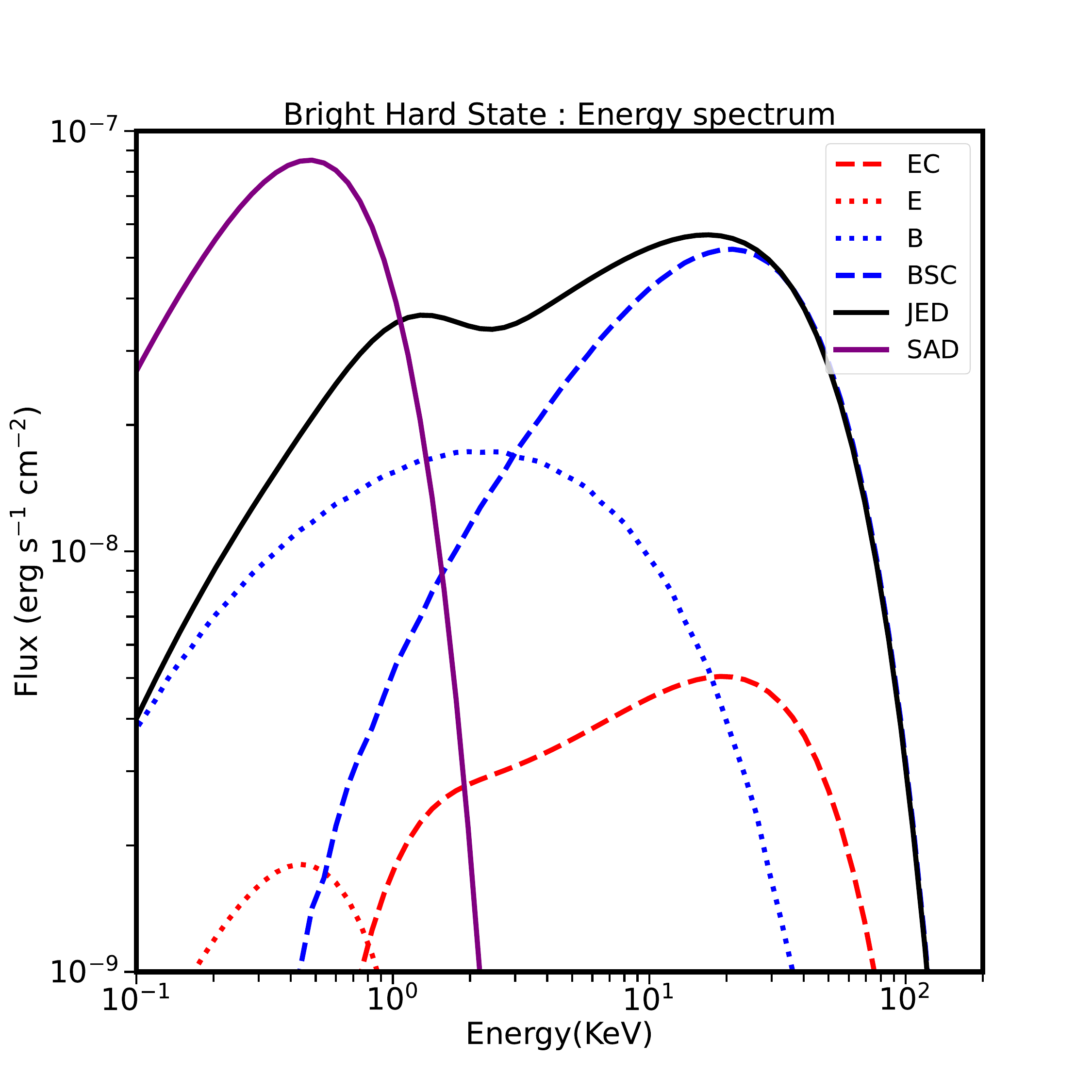}
 \caption{SED of the different radiative components of the JED--SAD best fit (using model \ref{model3}) of the bright hard state. The solid black curve represents the total JED component, the solid purple curve the SAD component, the dashed blue curve the Comptonised bremsstrahlung component, the dotted blue curve the bremsstrahlung component, and the dashed red curve the external photon Comptonised in the JED.
 \label{fig:bright_dyplo}}
\end{figure}

We report in Fig. \ref{fig:bright_dyplo} the total JED--SAD SED corresponding to the best fit of the bright hard state spectrum as well as the different spectral components, i.e. the  bremsstrahlung (B), self-Comptonised bremsstrahlung (BSC), synchrotron (S), self-Comptonised synchrotron (SSC), the external photon field (E) corresponding to photons emitted by the SAD and intercepted by the JED, parametrised by $\omega$ (see Sect. \ref{jedsadsect}), and its Comptonisation (EC). Our modelling suggests that, above $\sim$ 1 keV, the bright hard state spectrum is dominated by the BSC component. This differs from interpretations in which inverse Compton scattering of thermal disk photons is assumed to dominate the hard X-ray emission; this process is included in our model through the E and EC components, but remains subdominant in our best-fit solution.

In this best-fit solution, the electron density in the JED remains relatively constant around $10^{19}$ cm$^{-3}$, while the JED vertical optical depth reaches $\sim$10, corresponding to an optically thick configuration. The radial profiles of the physical quantities corresponding to this solution are reported (in blue) in Fig.~\ref{fig:diskstruc}. As discussed in the next section, this has important consequences for the polarisation properties.

\subsubsection{X-ray polarisation}
\label{Sec:Bright_polar}
Using the density, magnetic field, and temperature profiles of the bright hard state JED--SAD best-fit solution, we computed the polarisation properties and the spectrum of each radiative component with MONK. We assumed an isotropic magnetic field distribution together with an inclination angle of $70^\circ$, which is close to the upper bound inferred by \citet{wood2024}. This choice represents a limiting case that maximises the predicted polarisation, as lower inclinations would produce weaker polarisation signatures. Since the synchrotron emission is negligible in the total JED--SAD best-fit SED (see Sect.~\ref{sec412} and Fig.~\ref{fig:bright_dyplo}), the magnetic field configuration has no impact on the predicted polarisation properties. The polarisation signatures are obtained by averaging five independent Monte Carlo simulations, and the error bars correspond to the propagated standard errors estimated from their dispersion.

\begin{figure}
\centering

\includegraphics[width=\columnwidth]{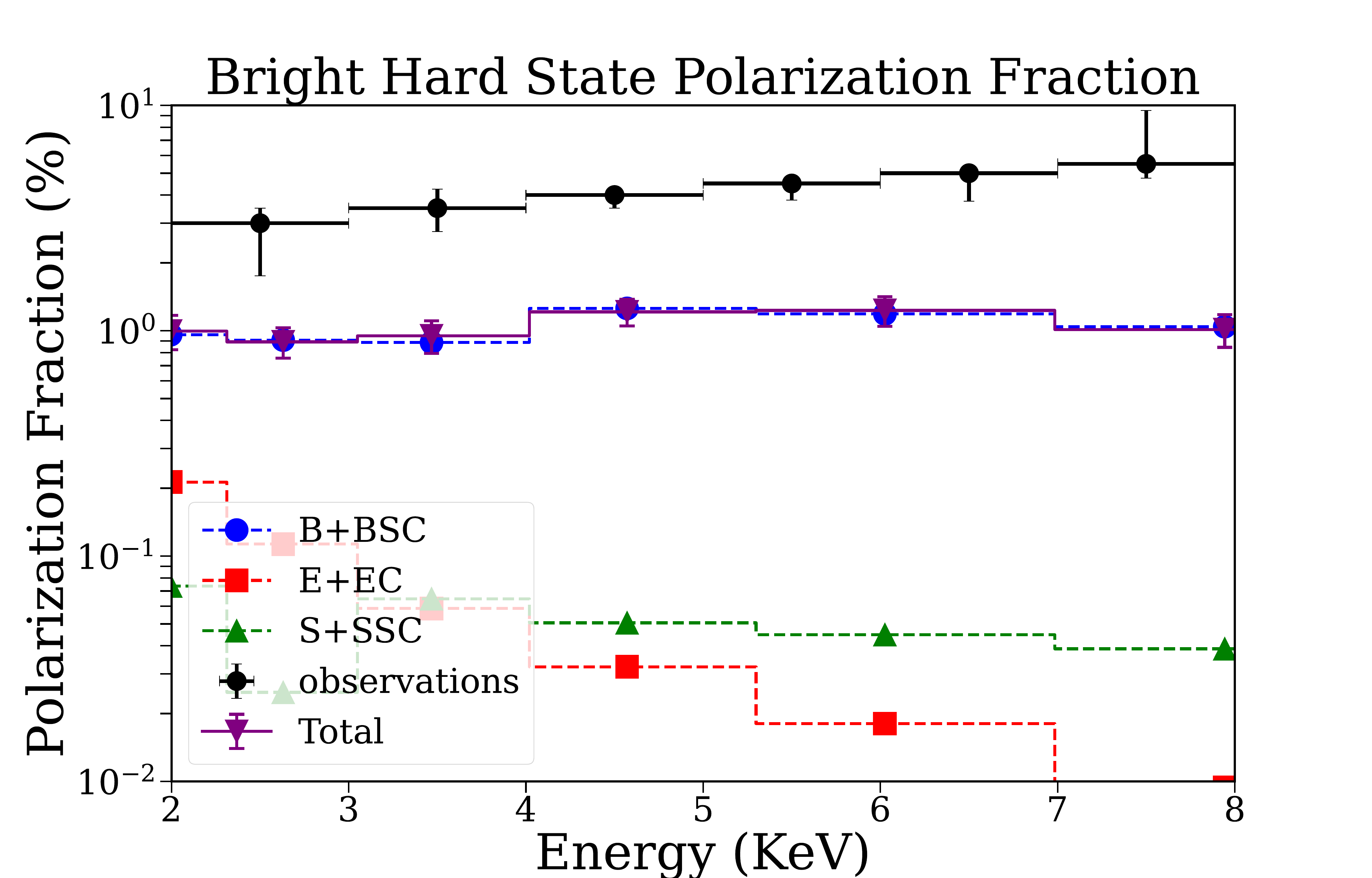}
\label{fig:Monk_bright_polfrac}

\includegraphics[width=\columnwidth]{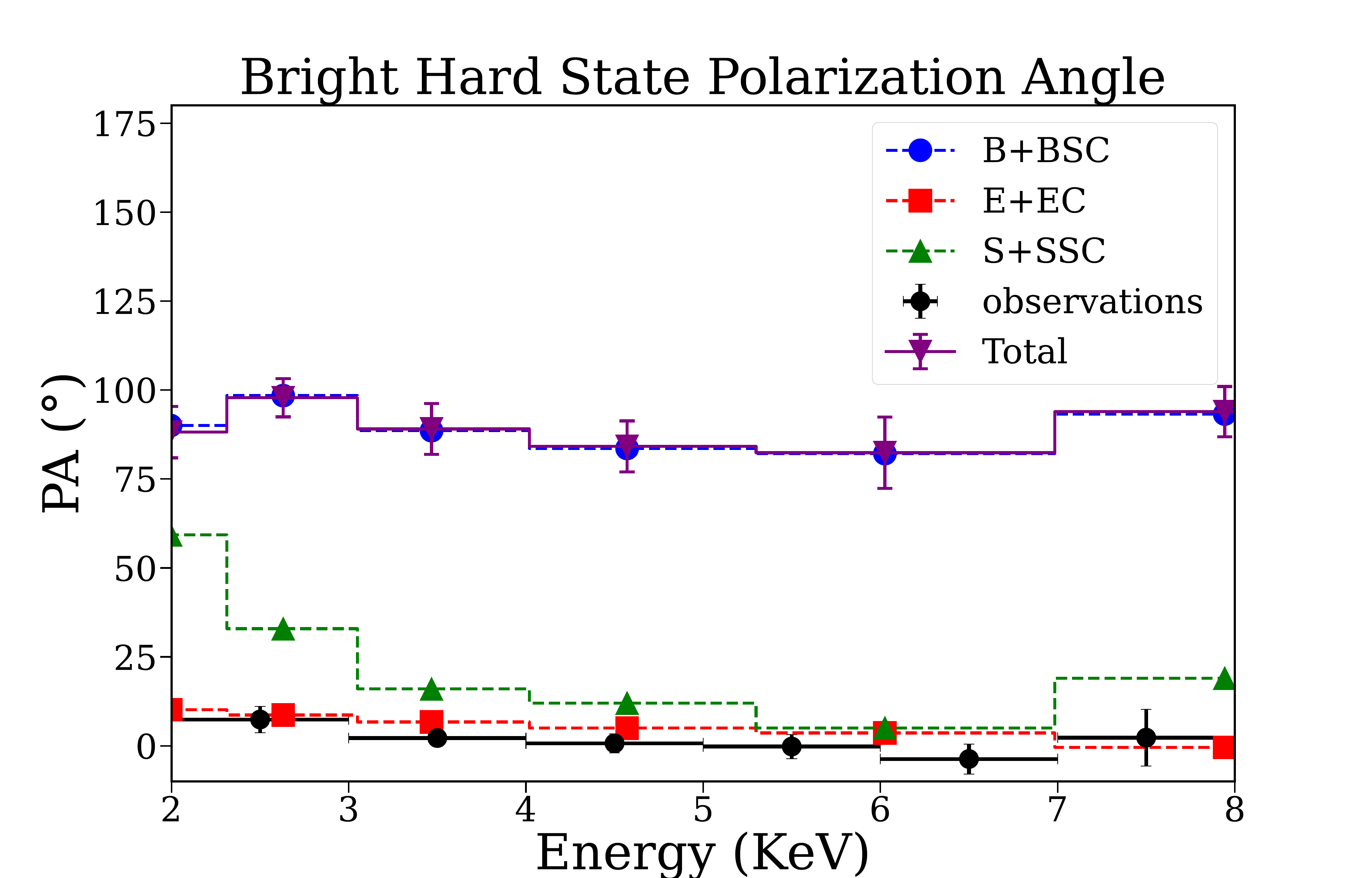}
\label{fig:Monk_bright_polang}

\caption{Observed \textit{IXPE} polarisation properties of the bright hard state of Swift J1727.8-1613 compared to the luminosity-weighted simulated polarisation signatures computed by MONK with our JED--SAD best-fit configuration. \textit{Top}: Polarisation fraction. \textit{Bottom}: Polarisation angle. In both panels, the \textit{IXPE} observations are in black, and we report the polarisation properties of each radiative process.}
\label{fig:pol_bright}
\end{figure}

Figure~\ref{fig:pol_bright} compares the observed \textit{IXPE} polarisation properties with the MONK predictions for the B+BSC, E+EC and S+SSC components together with the total JED--SAD emission. Each component is normalised by its own luminosity to facilitate comparison (hereafter referred to as luminosity-weighted). Since our best-fit bright hard state SED is dominated by the B+BSC component, the predicted total polarisation is almost identical to that of the B+BSC component.

The total simulated polarisation fraction is about 1\%, arising solely from the Comptonisation of thermal bremsstrahlung, whose seed photons are expected to be unpolarised. This is significantly below the observed polarisation fraction of 3--5\%. The polarisation angle is found to be $\sim90^\circ$ across the \textit{IXPE} energy range, in contrast with the observed $0^\circ$, which corresponds, in our reference frame, to a polarisation perpendicular to the accretion disk plane. This is surprising since, in our model, the emission is dominated by Comptonised bremsstrahlung and, given the JED geometry aligned with the accretion flow, we would have expected the polarisation angle to be perpendicular to the JED plane (see e.g. \citealt{krawczynski2022}). However, the large JED vertical optical depth ($\tau\sim10$) implies that the escaping photons mainly originate from the JED surface. They can propagate more freely in the vertical direction than horizontally, creating a preferred propagation direction and leading to a polarisation angle predominantly parallel to the disk plane. Reducing the optical depth by a factor of 5 ($\tau\sim2$) recovers the expected polarisation angle perpendicular to the disk plane, in agreement with \citet{dovciak2008, tomaru2024,Sunyaev1985}. We also note that the \textsc{simplcut} component is not included in our bright hard state polarisation simulations. As it describes the Comptonisation of disk photons in the warm corona, we expect its polarisation orientation to be consistent with that of the BSC emission. Its inclusion is therefore not expected to significantly affect our conclusions.

\subsection{Dim hard state}

\subsubsection{Spectral fit}

\begin{figure}
\includegraphics[width=\columnwidth]{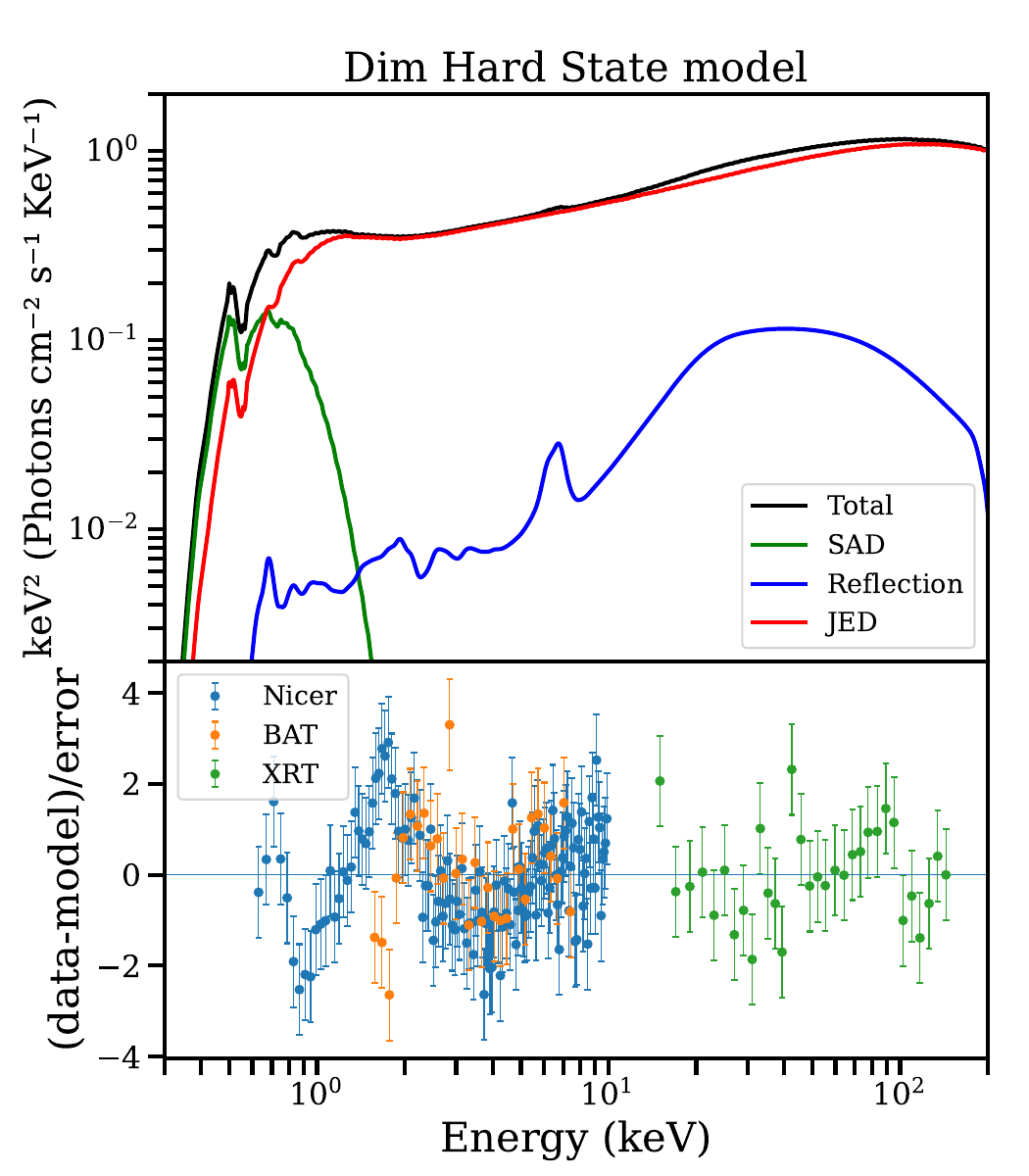}
 \caption{Dim hard state fit. \textit{Top}: $EF_E$ model with the SAD (green), JED (red), and reflection (blue) components. \textit{Bottom}: Residuals corresponding to \textit{NICER} (blue), \textit{Swift}/BAT (green), and \textit{Swift}/XRT (orange) data.\label{fig:dim_model}}
\end{figure}

The dim hard state was observed by \textit{IXPE}, \textit{NICER}, and \textit{Swift} (XRT and BAT). 
We used the same models as for the bright hard state (Sect. \ref{Sec:BrightHardState}) to fit these data. We first used the model given by Eq. (\ref{model1}), fixing the JED--SAD normalisation to 16.8 to be consistent with the value obtained for the bright hard state fits. Figure \ref{fig:dim_model} shows the best-fit SED obtained with this model as well as the residual. The model provides a satisfactory fit to the data, with a fit statistic of $\chi^2$= 193 for 193 degrees of freedom. Fitting this dataset with model \ref{model2}, while fixing the hard tail photon index and cutoff to the values obtained in the bright hard state fit, results in a hard tail normalisation that is consistent with zero. Additionally, including a simplcut component in model~\ref{model1} gives a best-fit value of the scattering fraction parameter consistent with zero, indicating that the data do not require this component. Model \ref{model1} appears then sufficient to give an acceptable fit of the dim hard state data. The best-fit parameters are listed in Table \ref{table:bestfitmodelparameter}, right column. The transition radius  $R_j=16.1 \pm 0.2\ Rg$ is broadly consistent with the value found for the bright hard state fit. The sonic Mach number $m_s=0.6 \pm 0.1$, slightly smaller than in the bright hard state. However, the accretion rate is lower by a factor of $\sim$10 with respect to the bright hard state with $\dot M=0.13 \pm 0.01$ $L_{Edd}/c^{2}$ rather than 1.6$\pm$0.1 $L_{Edd}/c^{2}$. Indeed, a lower accretion rate is expected given the two-order-of-magnitude difference in X-ray flux between the two hard states. The two-order-of-magnitude luminosity difference can arise from a $\sim10$ decrease in $\dot{M}$, since the bremsstrahlung emission dominating the bright hard state scales approximately as $\dot{M}^2$, while the synchrotron emission dominating the dim hard state scales approximately as $\dot{M}$ (see Fig. 9 in \citealt{marcel2022}).

\subsubsection{Radiative properties of the jet-emitting disk solution}

\begin{figure}
\includegraphics[width=\columnwidth]{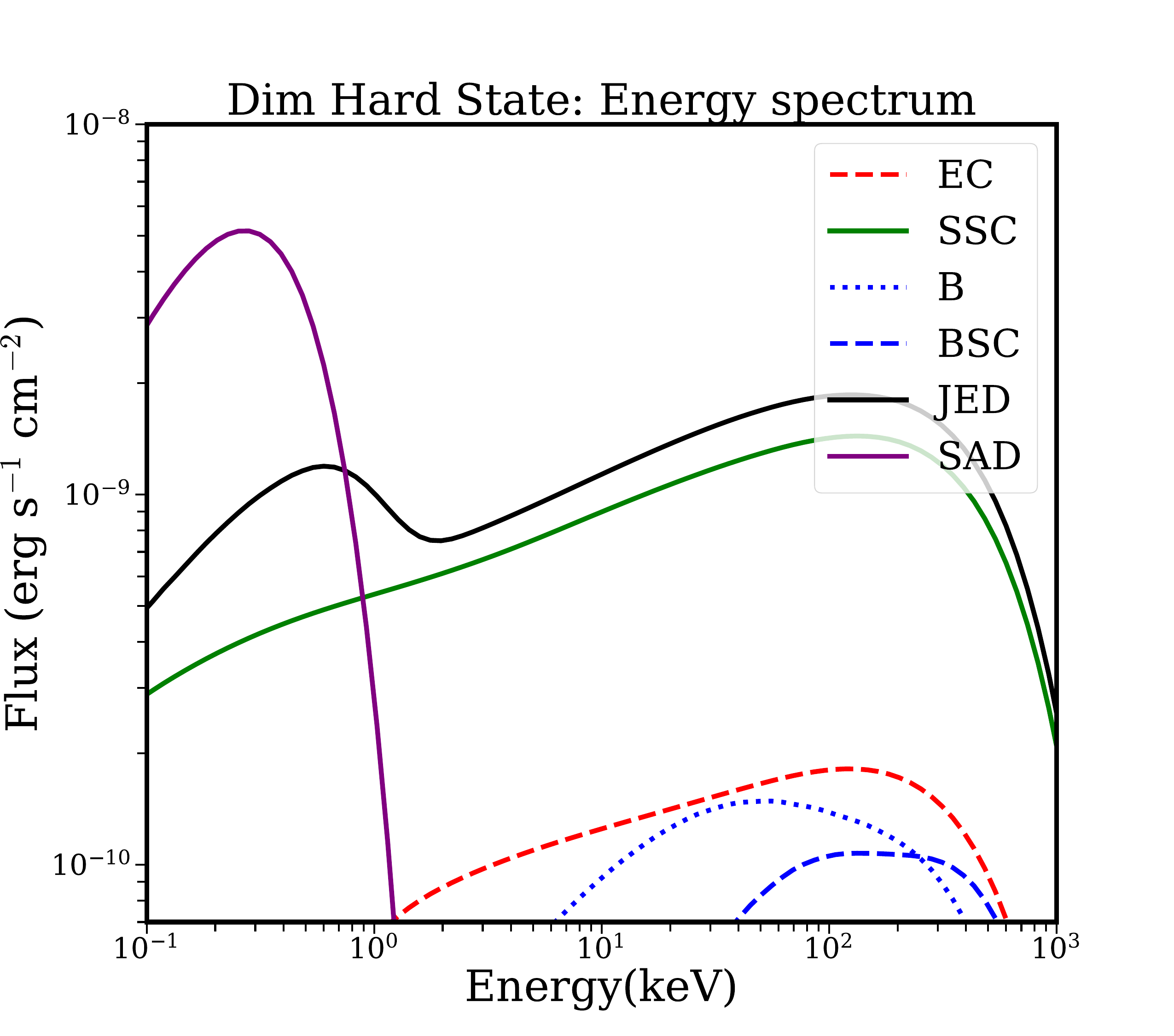}
 \caption{SED of the different radiative components of the JED--SAD best fit (using model \ref{model1}) of the dim hard state. The solid green curve represents the SSC component, the dotted blue curve the B component, the dashed blue curve the BSC component, the dashed red curve the EC component, the solid black curve the sum of these components, and the solid purple curve the SAD emission. \label{fig:dim_dyplo}}
\end{figure}

We report in Fig. \ref{fig:dim_dyplo} the total JED--SAD SED corresponding to the best fit of the dim hard state spectrum as well as the different spectral components. Our best-fit dim hard state solution is dominated by the synchrotron component. Within our modelling, the lower accretion rate inferred for the dim hard state with respect to the bright hard state leads to a switch in the dominant process from Comptonised bremsstrahlung to Comptonised synchrotron, due to the significant decrease in the density (the bremsstrahlung emissivity scales with density squared, while the synchrotron is proportional to the density). Within our model, this decrease is explained by the lower accretion rate derived for the dim solution, which is also around 10 times smaller. Meanwhile, the JED vertical optical depth is of the order of unity, about one order of magnitude lower than in the bright hard state, while the disk aspect ratio remains similar. The corresponding radial profiles are shown in Fig.~\ref{fig:diskstruc}.

\subsubsection{X-ray polarisation}

\begin{figure}
\centering

\includegraphics[width=\columnwidth]{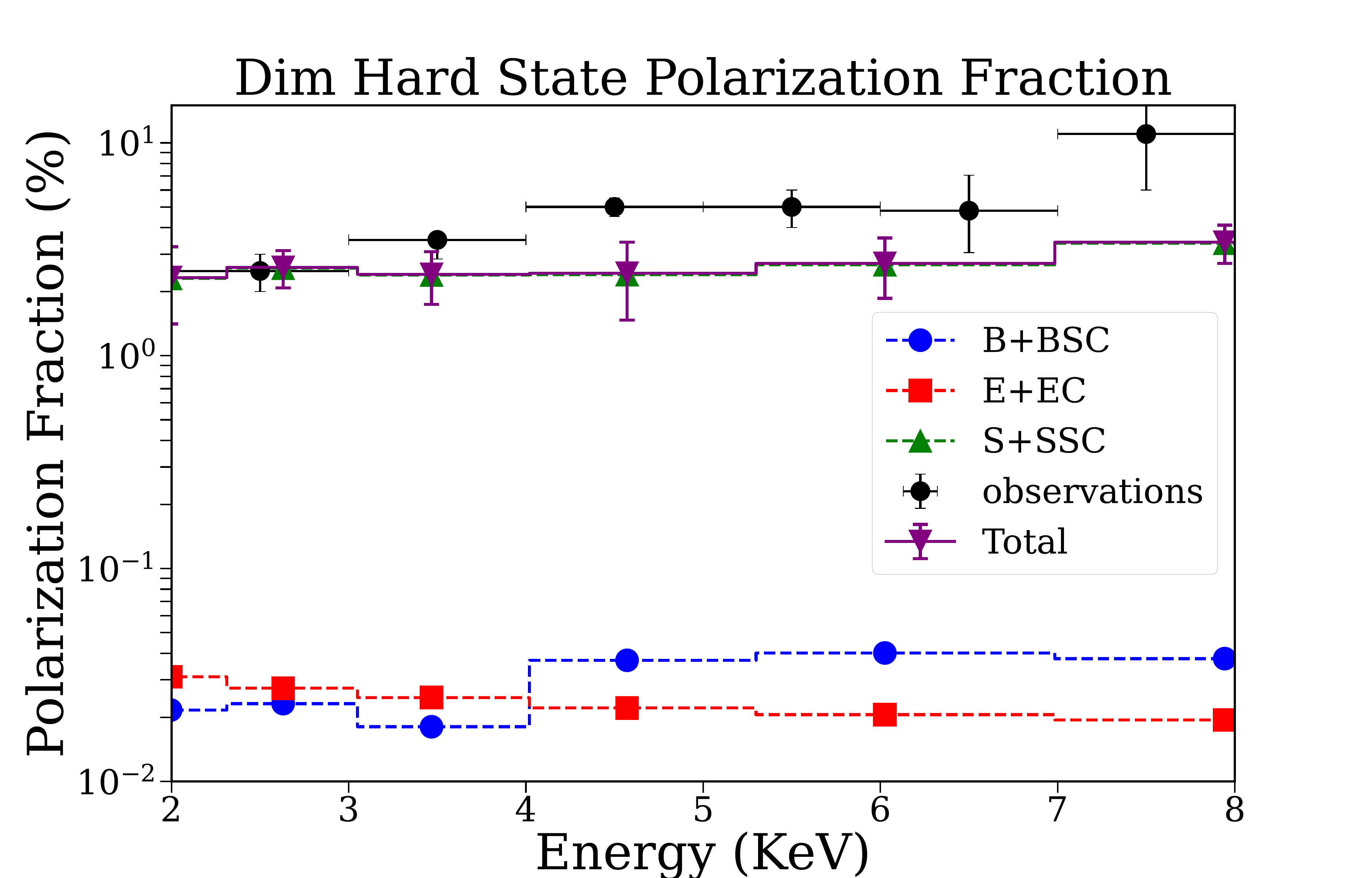}
\label{fig:Monk_bright_polfrac}

\includegraphics[width=\columnwidth]{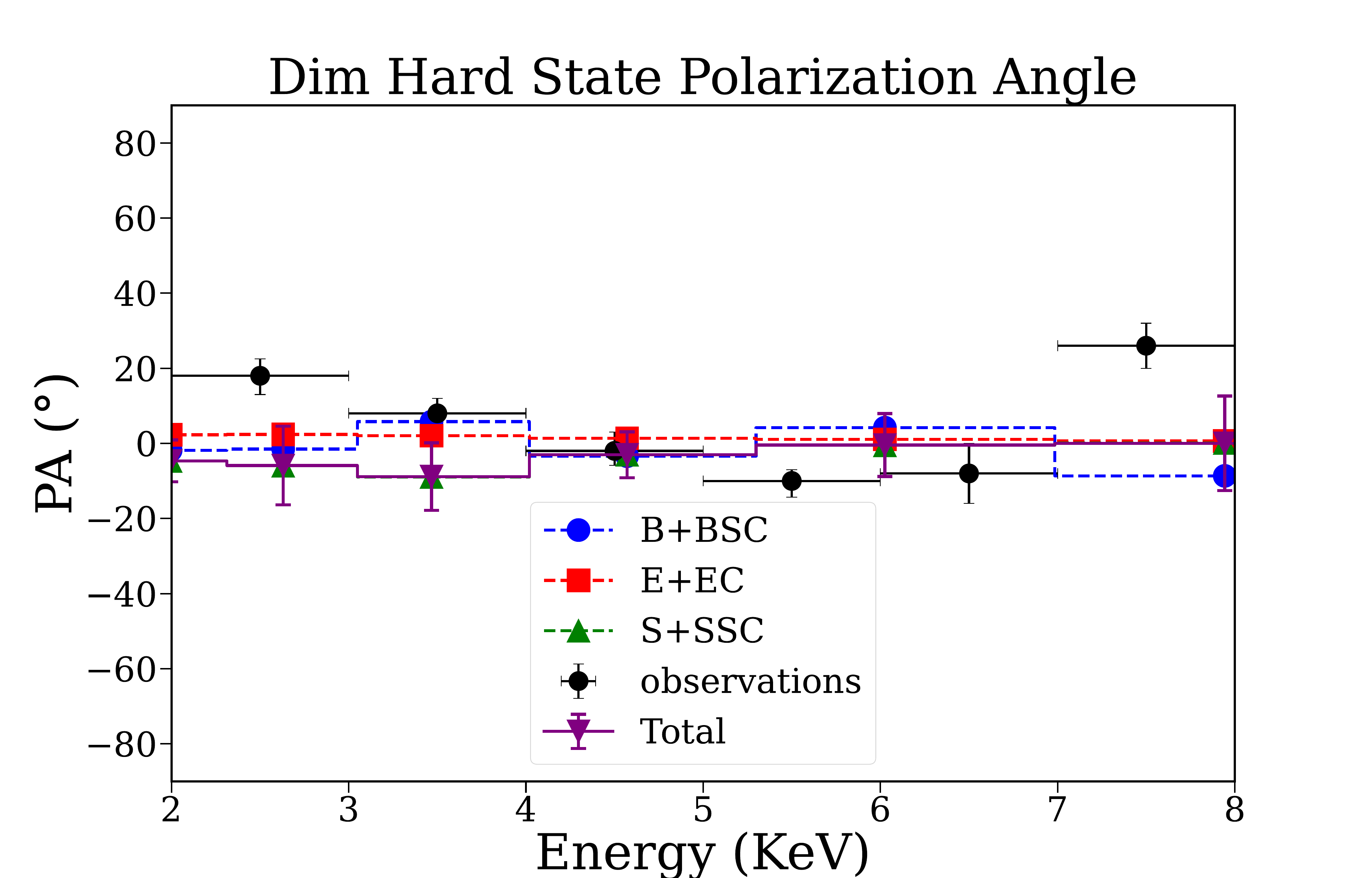}
\label{fig:Monk_bright_polang}

\caption{Observed and simulated (luminosity-weighted) polarisation signatures for the dim hard state JED--SAD configuration assuming a tangled magnetic field with no preferred orientation. \textit{Top}: Polarisation fraction. \textit{Bottom}: Polarisation angle. The \textit{IXPE} observation is shown as black points. In both panels we report the polarisation properties of each radiative process.}
\label{fig:pol_dim}
\end{figure}

Similarly to the bright hard state (see Sect.~\ref{Sec:Bright_polar}), we computed the polarisation properties of the dim hard state JED--SAD best-fit solution with MONK using the corresponding density, magnetic field, and temperature profiles. Since the JED emission in our best-fit solution is dominated by the SSC component, its polarisation properties could depend on the magnetic field configuration. We first assumed a tangled magnetic field with no preferred orientation (isotropic). Figure~\ref{fig:pol_dim} shows the simulated polarisation signatures using the same colour code as in Fig.~\ref{fig:pol_bright}. The simulated polarisation properties reproduce the observations reasonably well, both in terms of polarisation fraction and polarisation angle, although the slight increase in the polarisation degree with energy suggested by the \textit{IXPE} data is not reproduced.

We then tested different magnetic field configurations, assuming organised geometries (e.g. vertical or toroidal), for which synchrotron emission is expected to be more strongly polarised. However, our simulations show that the impact on the resulting polarisation is negligible because of the strong Faraday rotation occurring in the hottest part of the JED, which produces most of the emission in the 1--10 keV energy range and depolarises the synchrotron seed photons (see Fig.~\ref{fig:pol_dim_b_config}). In our simulations, the SSC polarisation is therefore entirely produced by Compton scattering.

\vspace{-0.25cm}
\section{Other components that could play a role in the polarisation signal}
\label{DiscSect}
While our JED--SAD spectral fits provide acceptable results for both hard states, the simulated polarisation reproduces the dim hard state only approximately and fails to reproduce the polarisation fraction and angle of the bright hard state. However, the MONK calculations include only the JED emission and neglect the polarisation associated with additional components, whether included in the spectral modelling or not. We therefore investigate their possible contribution to the simulated polarisation signatures in the following sections.

\subsection{Reflection component}
\label{Refl_sec}

Figures \ref{fig:bright_model} and \ref{fig:dim_model} show that part of the observed X-ray flux in the 2--8 keV \textit{IXPE} range comes from the reflection component. As a first simple test, we assumed a constant intrinsic reflection polarisation fraction (RPF) of 5\% or 15\%, with a reflection polarisation angle (RPA) perpendicular to the disk plane. This is in rough agreement with lamp-post simulations (e.g. \citealt{matt1993,podgorny2023}), where even larger intrinsic polarisation ($\sim$10--20\%) is expected for the $70^\circ$ inclination assumed for Swift J1727.8-1613.

\begin{figure}
\centering

\includegraphics[width=\columnwidth]{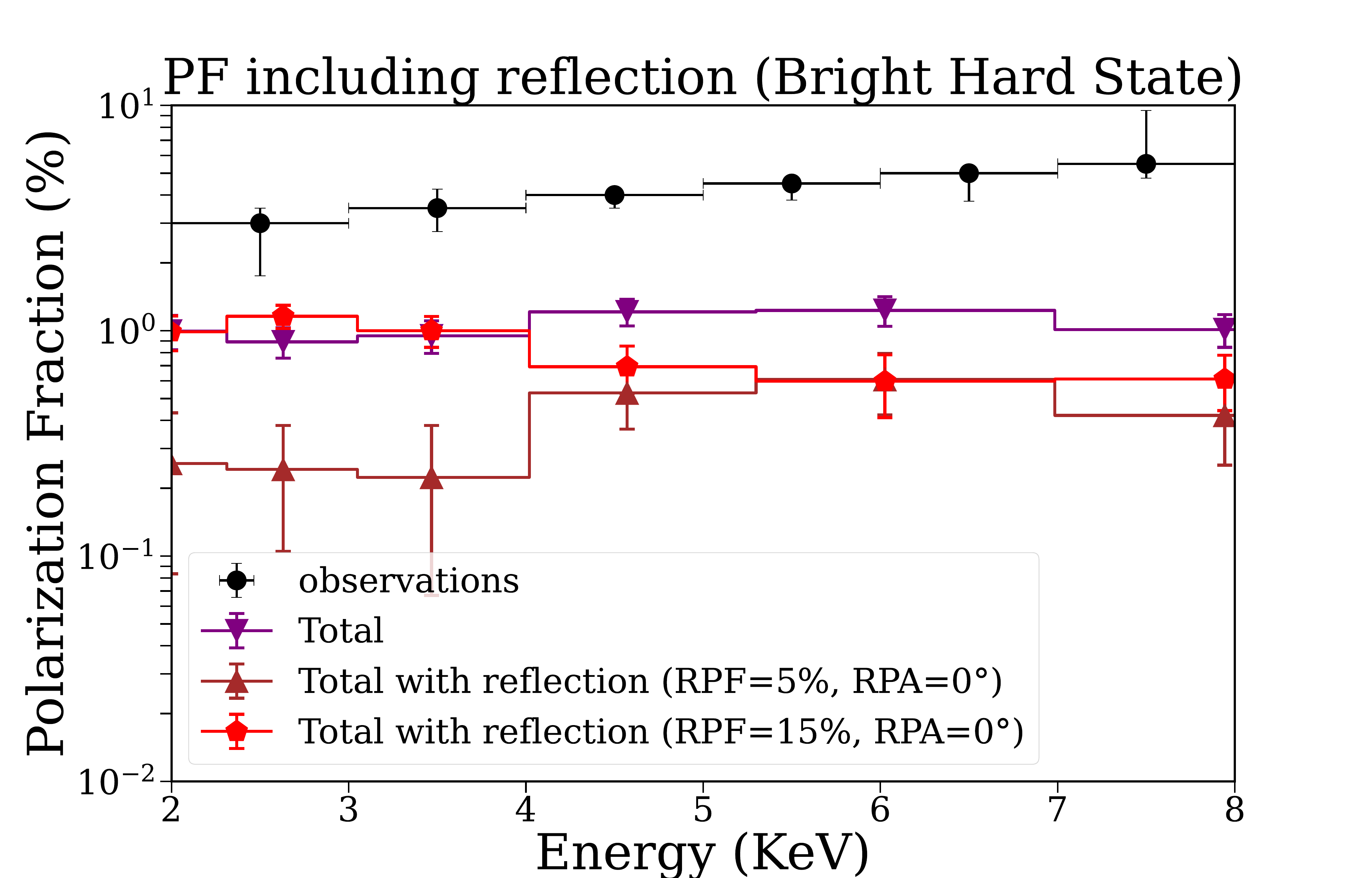}
\label{fig:Monk_bright_polfrac_refl}

\includegraphics[width=\columnwidth]{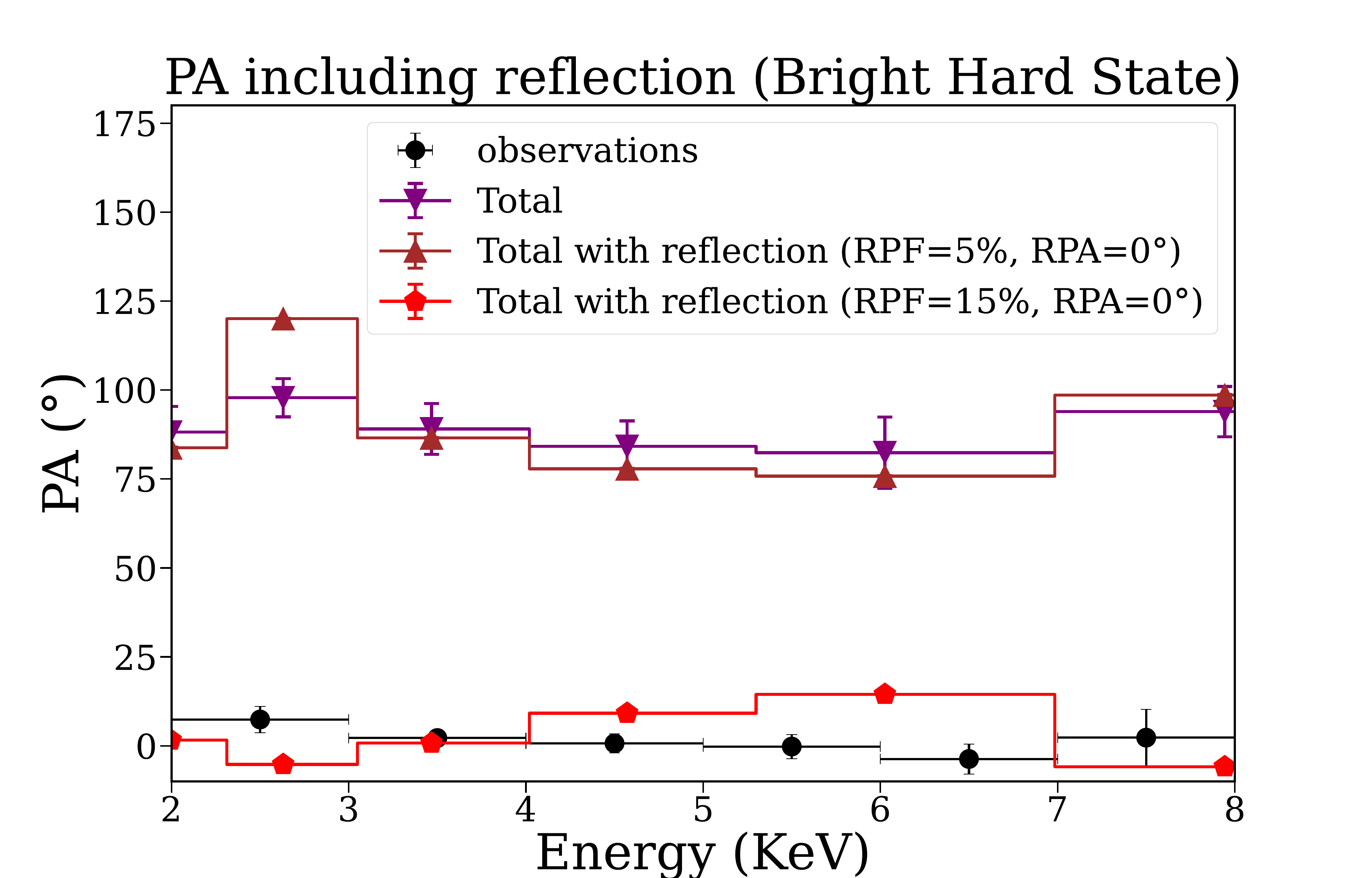}
\label{fig:Monk_bright_polang_refl}

\caption{Observed and simulated (luminosity-weighted) polarisation signatures for the bright hard state JED--SAD configuration. \textit{Top}: Polarisation fraction. \textit{Bottom}: Polarisation angle. The \textit{IXPE} observation is shown as black points. In both panels we report the simulated polarisation properties obtained with MONK with different supposed RPFs.}
\label{fig:pol_bright_refl}
\end{figure}

In the bright hard state, the addition of reflection affects the total polarisation differently depending on the RPF (see Fig. \ref{fig:pol_bright_refl}). Since the JED and reflection components have different polarisation angles, their combination through the Stokes parameters can modify both the polarisation fraction and its energy dependence. For RPF = 5\%, the total emission remains weakly polarised ($<0.5\%$). For RPF = 15\%, the polarisation angle becomes perpendicular to the disk, as observed by \textit{IXPE}, but the total polarisation fraction remains around 1\%, significantly below the observed value. This is expected since the reflection flux is approximately one order of magnitude lower than the total flux (see Fig.~\ref{fig:bright_model}).

\begin{figure}
\centering

\includegraphics[width=\columnwidth]{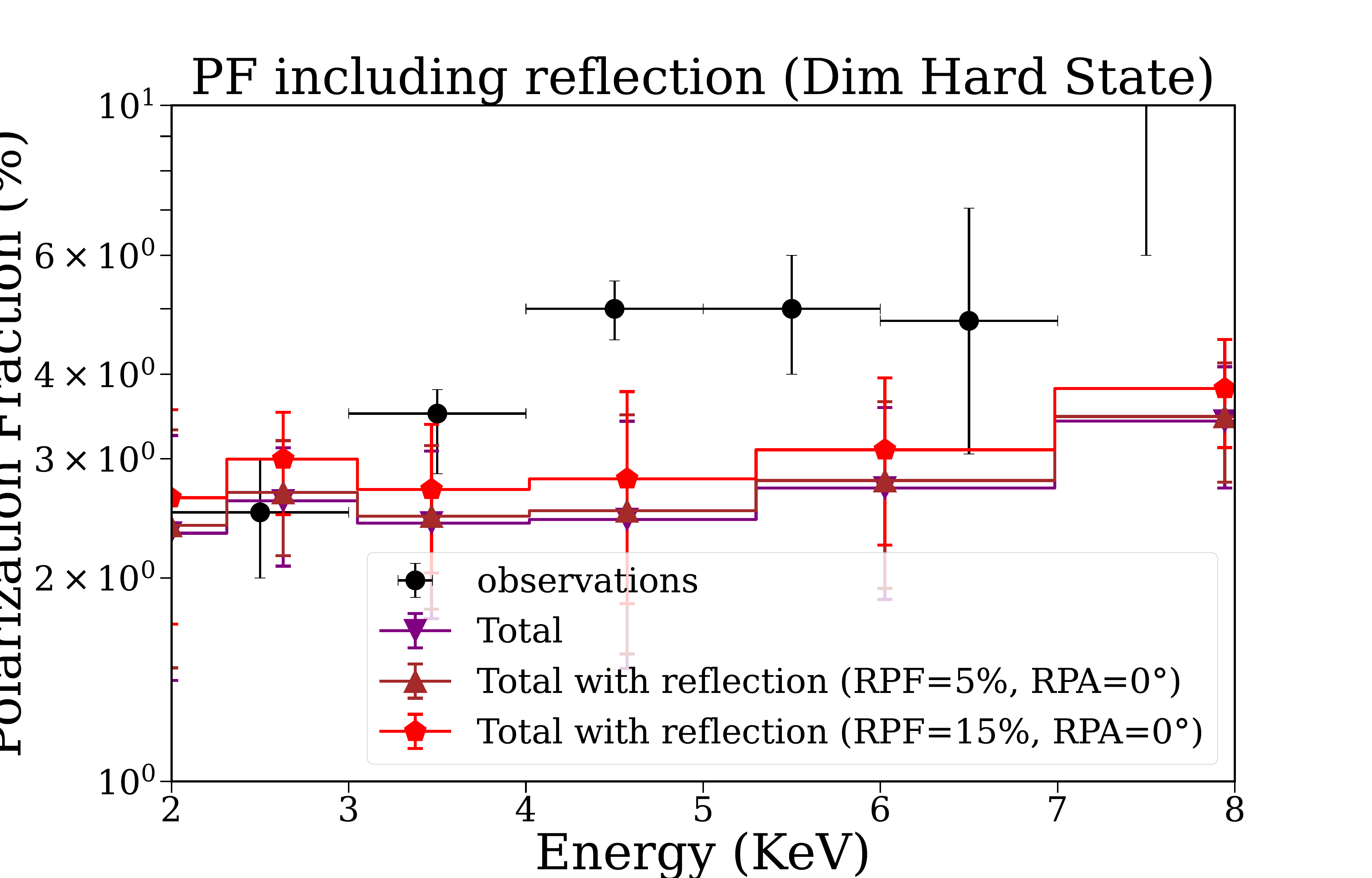}
\label{fig:Monk_dim_polfrac_refl}

\includegraphics[width=\columnwidth]{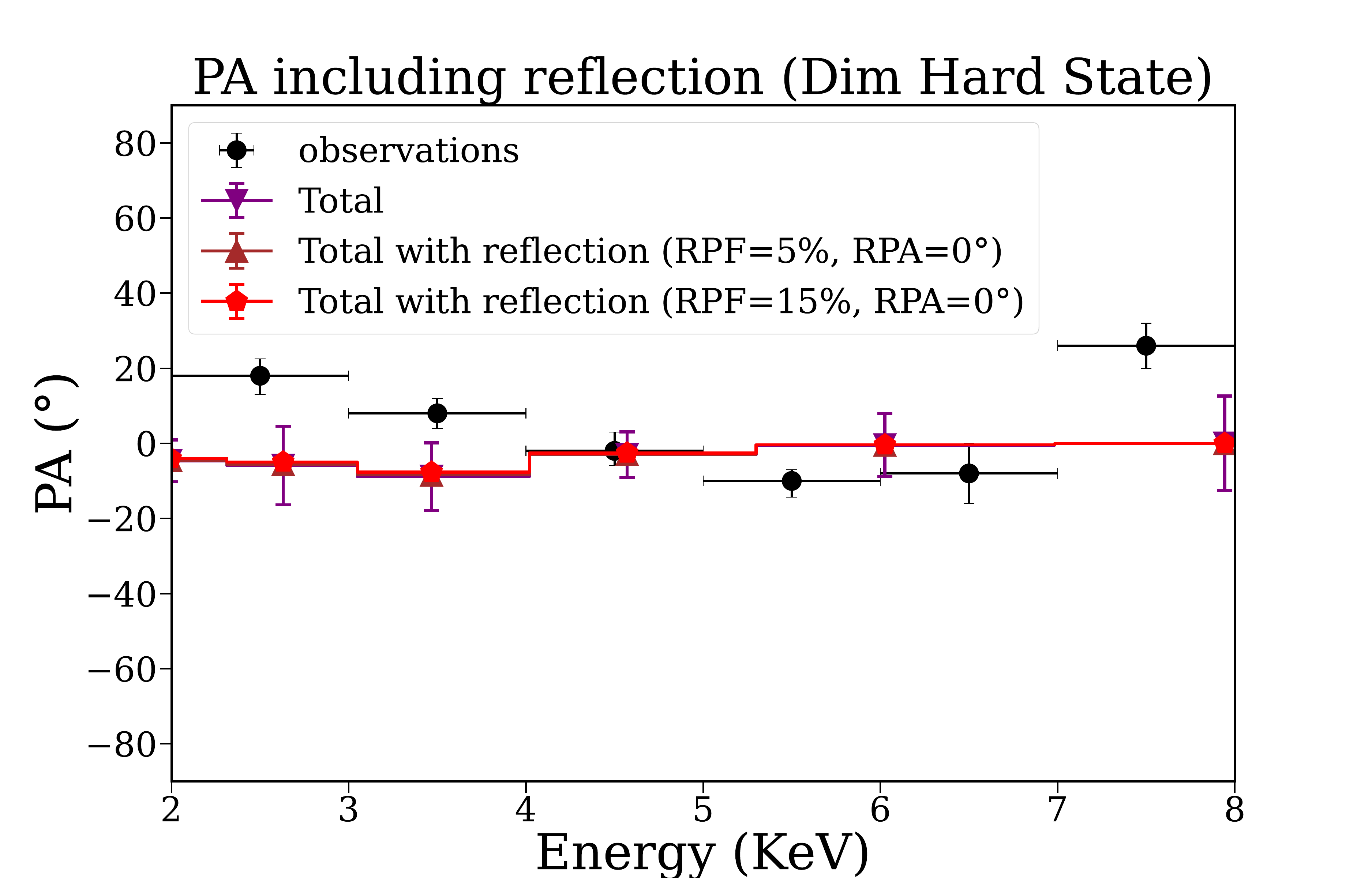}
\label{fig:Monk_dim_polang_refl}

\caption{Similar to Fig. \ref{fig:pol_bright_refl} but for the  dim hard state JED--SAD configuration.  Purple shows the total polarisation signature, brown the total polarisation signature including a RPF=5\% and RPA=$0^\circ$, and red the total polarisation signature including a RPF=10\% and RPA=$0^\circ$.}
\label{fig:pol_dim_refl}
\end{figure}

In the dim hard state, the JED and reflection components have the same polarisation angle, so the addition of reflection increases the total polarisation fraction (see Fig. \ref{fig:pol_dim_refl}). However, since the reflection flux is two orders of magnitude lower than the total flux (see Fig. \ref{fig:dim_model}), an RPF of 5--15\% does not significantly change our results. Overall, the polarisation brought by the reflection component is therefore not sufficient to reproduce the observations.

\subsection{Hard tail}
\label{Sec:nth}

The presence of a hard tail in the bright hard state suggests the presence of non-thermal particles \citep{Chand2026}, either within the JED or in a distinct component (e.g. the jet). The current JED--SAD/MONK framework assumes a purely thermal particle distribution, although a hybrid thermal/non-thermal distribution can be expected \citep{ghisellini1998,coppi1999,vurm2008,malzac2008}. Non-thermal particles can produce more polarised emission because of anisotropies (in velocity and energy) associated with the acceleration process; for example, pure non-thermal synchrotron emission in an ordered magnetic field can reach $\sim70\%$ polarisation \citep{rybicki1979_radiative_processes}. A lower polarisation is expected for a hybrid distribution due to Compton scattering, Faraday rotation, and dilution by the less polarised thermal emission \citep{beheshtipour2017}. Regardless of the origin of these non-thermal particles, we assess in this section the degree of polarisation required from the hard tail to reproduce the bright hard state polarisation features.

\begin{figure}
\centering

\includegraphics[width=\columnwidth]{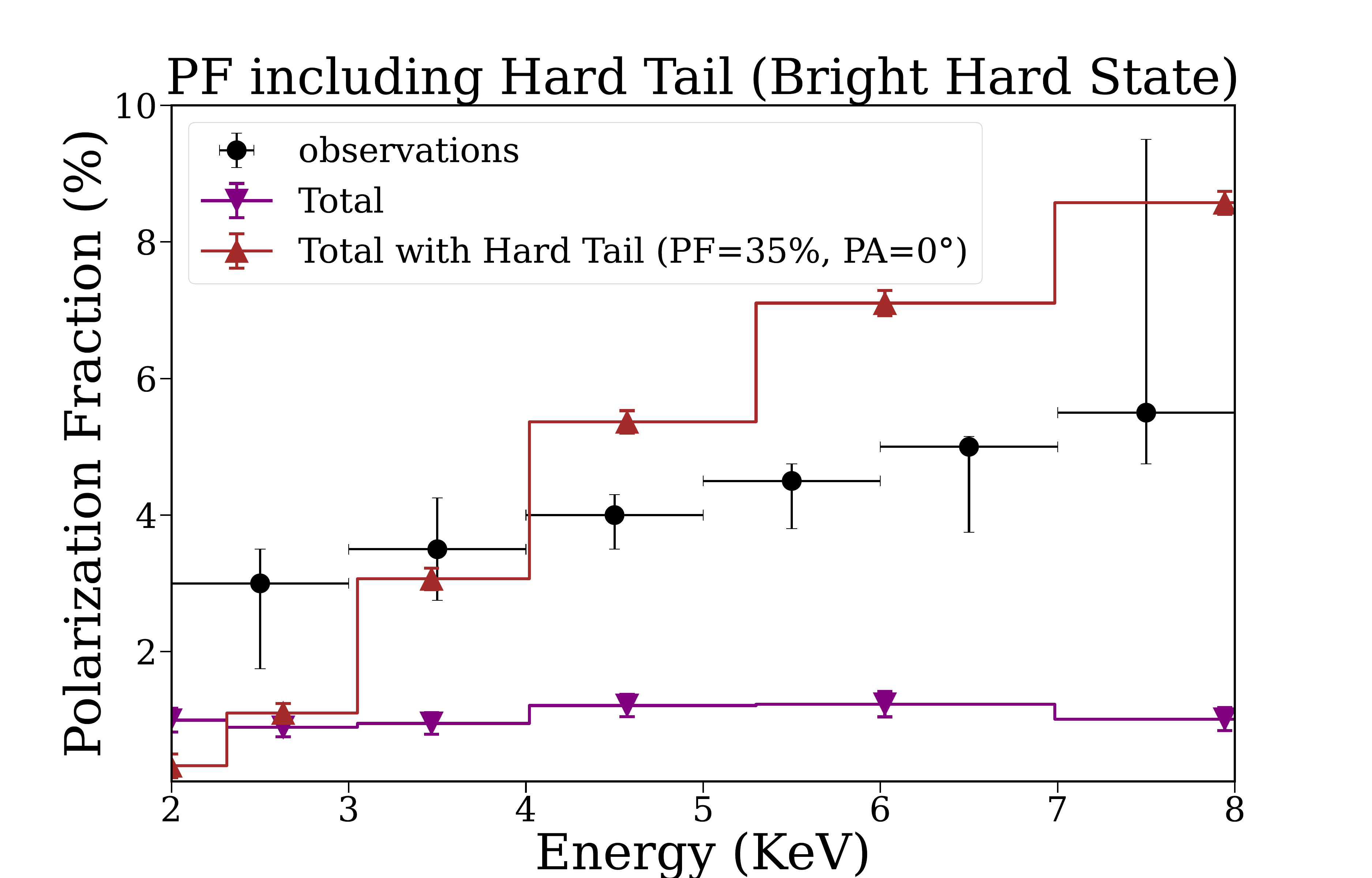}
\label{fig:Monk_bright_polfrac_ht}

\includegraphics[width=\columnwidth]{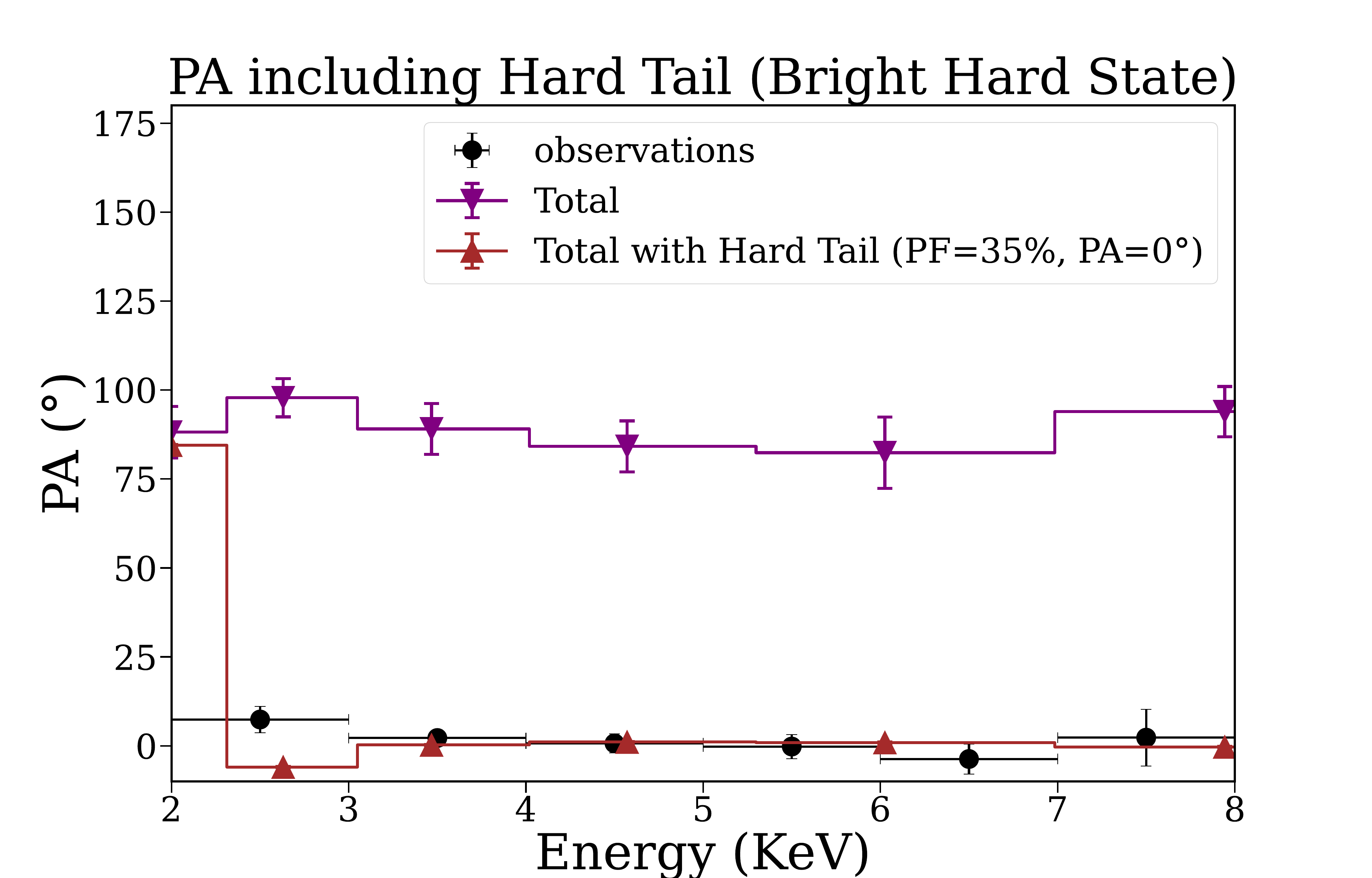}
\label{fig:Monk_bright_polang_ht}

\caption{Observed and simulated (luminosity-weighted) polarisation signatures for the bright hard state JED--SAD configuration. \textit{Top}: Polarisation fraction. \textit{Bottom}: Polarisation angle. The \textit{IXPE} observation is shown as black points. In both panels we report the polarisation properties of MONK, including the hard tail component computed by hand.}
\label{fig:pol_bright_ht}
\end{figure}

As discussed in Sect. \ref{Sec:Bright_polar},  we estimated the JED polarisation in this state to be $\sim$1\% and
oriented parallel to the disk surface. To obtain, with the addition of the hard tail, a polarisation of $\sim$3-5\% perpendicular to the disk (as observed), the hard
tail needs to be significantly polarised, with a polarisation angle perpendicular
to the disk. Given its flux in the \textit{IXPE} energy range, a hard-tail polarisation fraction of $\sim35\%$ is the minimum required to reproduce the observations. This remains compatible with the upper limit reported above 100 keV by \cite{bouchet2024}, provided that the hard-tail polarisation does not vary significantly with energy. As shown in Fig.~\ref{fig:pol_bright_ht}, its addition reproduces the data at first order. Below 3 keV the hard-tail flux becomes negligible due to the low energy cutoff (expabs) apply to this component. The JED then becomes dominant, which explains the change in polarisation signatures (see Fig.~\ref{fig:bright_model}). 

\subsection{Winds}
\label{Sec:winds}
Another mechanism that may contribute to the observed polarisation is scattering in the disk wind. Electron scattering in ionised disk winds can produce polarisation degrees of a few percent, depending on the wind geometry, optical depth, and viewing inclination, and can therefore modify both the polarisation fraction and angle \citep{Nitindala2025, matt1993,tomaru2024}. Observational support comes from phase-resolved \textit{IXPE} observations of eclipsing neutron-star X-ray binaries, where variations in the polarisation properties have naturally been interpreted as changes in the contribution of scattered radiation from an extended accretion-disk atmosphere and/or disk wind \citep{Mikusincova2025, Ursini2024}. Although Swift J1727.8$-$1613 is not an eclipsing system, these results show that scattering in extended outflows can significantly alter the observed X-ray polarisation and should therefore be considered as a possible contribution to its polarisation signal.

\section{Energy budget, caveats, and comparisons}
\label{Sec:Energy_budget_caveats}
As a physically motivated model, the JED--SAD framework is subject to a well-defined energy budget. In the bright hard state, the additional warm-corona and hard-tail components may significantly modify this budget. For the JED region ($R_{\rm ISCO}<R<R_J$), the JED$-$SAD energy budget can be expressed as\vspace{-0.15cm}
\begin{equation}
P_{\rm acc}^{\rm JED} = 2P_{\mathrm{jet}} + P_{\mathrm{rad}} + P_{\mathrm{adv}},
\end{equation}
where $P_{\mathrm{jet}}$ is the power carried away by one of the two jets, produced at each side of the magnetised accretion flow,  $P_{\mathrm{rad}}$ is the radiative power emitted by the disk, and $P_{\mathrm{adv}}$ is the power advected inwards with the accretion flow \citep{marcel_unified_2018}. The jet flux is 
\vspace{-0.15cm}
\begin{equation}
2P_{\mathrm{jet}} = b \times P_{\rm acc}^{\rm JED} = 8.64 \times 10^{-8}\ \mathrm{erg\ cm^{-2}\ s^{-1}},
\end{equation}
where $b=0.3$ is the jet power fraction (see Sect. \ref{jedsadsect}). To assess the impact of the additional spectral components on this energy budget, we compared their contributions to the observed flux in different energy bands. 

In the 0.01--10 keV energy band, the Simplcut component adds a flux of $9.17\times10^{-8}$ erg cm$^{-2}$ s$^{-1}$ to the SAD, corresponding to approximately 30\% of the total flux. In our spectral modelling, Simplcut represents a warm Comptonisation layer above the SAD. If this warm corona is powered by the accretion flow below $R_J$, this emission should be included in the available accretion power, requiring mainly a redistribution between thermal disk and Comptonised emission rather than a modification of the global energy budget. The resulting change on the thermal emission may then require an increase in $\dot M$ to fit the data.

In the 10--150 keV energy band, the hard-tail component has a flux of $4.78 \times 10^{-8}\ \mathrm{erg\ cm^{-2}\ s^{-1}}$, accounting for about 40\% of the total flux in this band and roughly half of $2P_{\mathrm{jet}}$. If this component originates from the jet, then a corresponding fraction of the jet power would be radiated instead of contributing to the radio-emitting outflow. Alternatively, if it is produced by a non-thermal particle population within the JED, it should be included in the radiative budget of the accretion flow. The similarity between $P_{\mathrm{jet}}$ and the hard-tail flux may therefore hint at a physical connection between the two (see Sect.~\ref{Sec:nth}). Neither the jet contribution nor the hard-tail emission is currently included self-consistently in the DYPLO spectral model. The integration of DYPLO with the ISHEM jet code \citep{malzac2013,malzac2014} is currently in progress.

Moreover, a limitation of the current JED$-$SAD framework concerns its geometrical description. The model is radially stratified, allowing the physical properties of the accretion flow to vary with radius, but it does not include any vertical stratification. Yet, our analysis indicates significantly different optical depths in the two states, with $\tau \sim 1$ in the dim hard state and $\tau \sim 10$ in the bright hard state. Such a large difference strongly suggests that the vertical structure of the flow plays an important role in shaping the emerging radiation field and, consequently, the polarisation properties predicted by the JED$-$SAD model. Interestingly, the simpler model of \citealt{podgorny2024} also requires a similar change in optical depth to reproduce the observations, showing that this requirement is not specific to the JED$-$SAD framework.

Our best-fit parameters can also be compared with previous studies of Swift J1727.8-1613. In particular, $R_J$ can be compared with the coronal radius $R_{\rm cor}$ inferred by \citet{ma2025}, as both characterise the radial extent of the inner hot flow. Our $R_J$ is smaller than $R_{\rm cor}$ in the bright hard state, but consistent with it in the dim hard state, although a quantitative comparison would require modelling the same Insight--HXMT (Hard X-ray Modulation Telescope) data within the JED--SAD framework. The JED temperature and optical depth can also be compared with the slab-geometry calculations of \citet{podgorny2024}. In the dim hard state, the relatively constant temperature ($\sim2.7\times10^9$ K and $kT_e\sim230$ keV) and optical depth ($\tau\sim0.46$) are broadly consistent with their calculations, whereas the strong temperature gradient in the bright hard state precludes such a direct comparison.

Finally, the combined use of DYPLO and MONK raises consistency issues because MONK accounts for non-local radiative effects, allowing photons produced at one radius to interact with the flow at other radii, whereas DYPLO relies on a local treatment. These effects should be more important in optically thin regimes and therefore in the dim hard state. Their precise impact on the polarisation remains difficult to assess, while their spectral effects are reported in Appendix~\ref{Appendix}.


\section{Conclusion}
We used the JED--SAD model to investigate the spectral and polarisation properties of two hard-state observations of the BHXB Swift J1727.8-1613. The bright hard state was observed at the top of the rising hard-state branch, at the beginning of the outburst, while the dim hard state was observed at the beginning of the decaying hard-state branch, at the end of the outburst. Although the these two observations differ by two orders of magnitude in X-ray flux, they show very similar X-ray polarisation properties as observed by \textit{IXPE}.

The final fits yield similar JED--SAD parameters for the two states, except for the accretion rate, which is about ten times lower in the dim hard state. While a reflection component is required in both cases, the bright hard state also requires a hard tail and a warm corona to achieve a satisfactory fit. Within our modelling, the lower accretion rate in the dim hard state leads to a change in the dominant radiative process, from bremsstrahlung and Comptonised bremsstrahlung in the bright hard state to synchrotron self-Compton emission in the dim hard state. The corresponding model predictions therefore produce different polarisation signatures, contrary to the \textit{IXPE} observations.

Our bright hard state simulations show a polarisation fraction of about 1\% in the 2--8 keV energy band. The simulated bright hard state polarisation angle is perpendicular to the jet axis, in disagreement with the observations, due to the high ($\sim$10) JED optical depth. In contrast, the polarisation properties predicted for the dim hard state are broadly consistent with, although slightly lower than, the observed ones.

To explain the disagreement in the bright hard state, we investigated whether part of the observed polarisation could originate from the reflection component, the hard tail, or winds. Even assuming a RPF of 15\% and a polarisation angle perpendicular to the disk, in agreement with lamp-post simulations (e.g. \citealt{matt1993,podgorny2023}), the total simulated polarisation fraction remains around 1\%, still significantly below the observed value. However, the polarisation angle becomes perpendicular to the disk, as observed. A more realistic treatment of the reflection is required to reach a firm conclusion. Concerning the hard tail, a polarisation fraction of about 35\%, with a polarisation angle parallel to the jet axis (consistent with the $>$100 keV polarisation measurements; \citealt{bouchet2024}), is required to reproduce the observed polarisation. The implementation of jet emission (e.g. the ISHEM model; \citealt{malzac2013,malzac2014}) and of a hybrid particle distribution in the JED--SAD framework will therefore be required to investigate the origin and polarisation properties of this hard tail. Moreover, scattering in an accretion-disk wind may also contribute to the observed polarisation \citep{matt1993,Nitindala2025,Mikusincova2025,Ursini2024}. This contribution could therefore modify the observed polarisation in addition to the intrinsic JED emission. Implementing radiative transfer through disk winds will be required to quantify its impact on the simulated polarisation signatures.

Finally, the current JED--SAD/MONK framework relies on a local treatment of radiative processes and does not account for non-local radiative transfer effects, which may become important in optically thin regions. A fully self-consistent coupling between the two frameworks will be the subject of future work.

\begin{acknowledgements}
We acknowledge support from SciServer and the High Energy Astrophysics Science Archive Research Center.

This work was supported by the French National Research Agency through the ``Investissements d'avenir'' program (ANR-15-IDEX-02), and partly by the CNRS National Programme for Astronomy and Astrophysics and the French space agency CNES.

Wenda Zhang acknowledges support from the National Natural Science Foundation of China (grants 12573019, 12333004, and 12433005) and the Strategic Priority Research Program of the Chinese Academy of Sciences (grant XDB0550201). Grégoire Marcel acknowledges support from the Polish National Science Center (grant 2023/48/Q/ST9/00138) and the Academy of Finland (grant 355672). Giorgio Matt acknowledges support from the Center of Theoretical Physics of Grenoble-Alpes (CPTGA) for a one-month visiting period at the Institute of Planetology and Astrophysics of Grenoble. Noémie Barnier acknowledges support from JSPS KAKENHI grants JP24K00673 and JP23H04892. Tristan Bouchet acknowledges support from the DFG/LIS project SI 2502/6-1 (project number 551127478).
\end{acknowledgements}

\bibliographystyle{aa}
\bibliography{references}

\begin{appendix}
\nolinenumbers

\section{Best fit details and analysis} \label{Appendix_bestfit}

\begin{figure}[H]
\includegraphics[width=0.90\columnwidth]{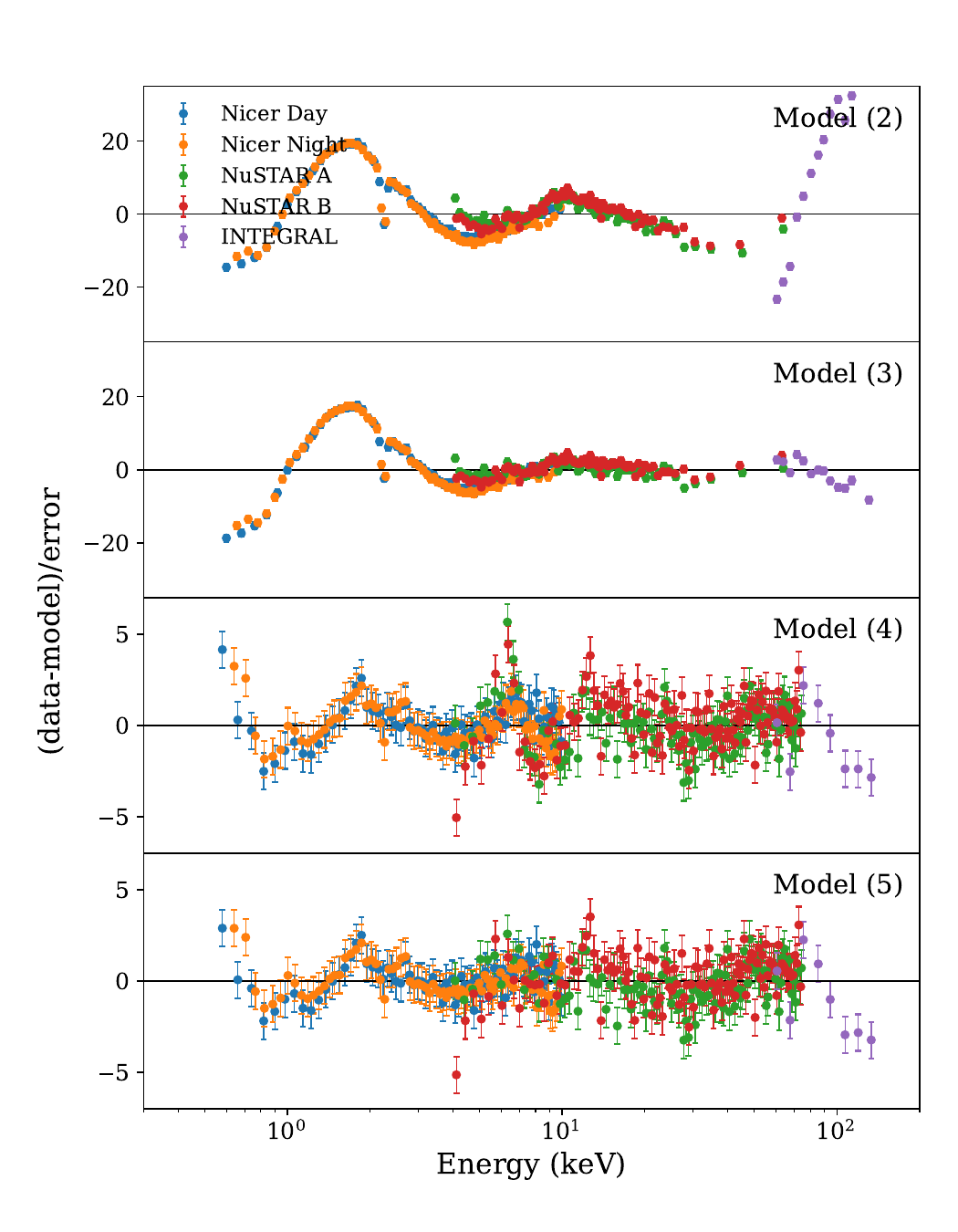}
 \caption{Bright hard state best-fit residuals obtained using \textit{NICER}, \textit{NuSTAR}, and \textit{INTEGRAL} data. We use \textit{NICER} day and night, \textit{NuSTAR} A and B, and \textit{INTEGRAL} data. {\it Top panel}: Model given by Eq. (\ref{model1}). {\it Second panel}: Model given by Eq. (\ref{model2}). {\it Third panel}: Model given by Eq. (\ref{model3_prev}). {\it Bottom panel}: Model given by Eq. (\ref{model3}).\label{fig:bright_preliminary_model}}
\end{figure}

\begin{figure}[H]
\centering

\includegraphics[width=0.75\columnwidth]{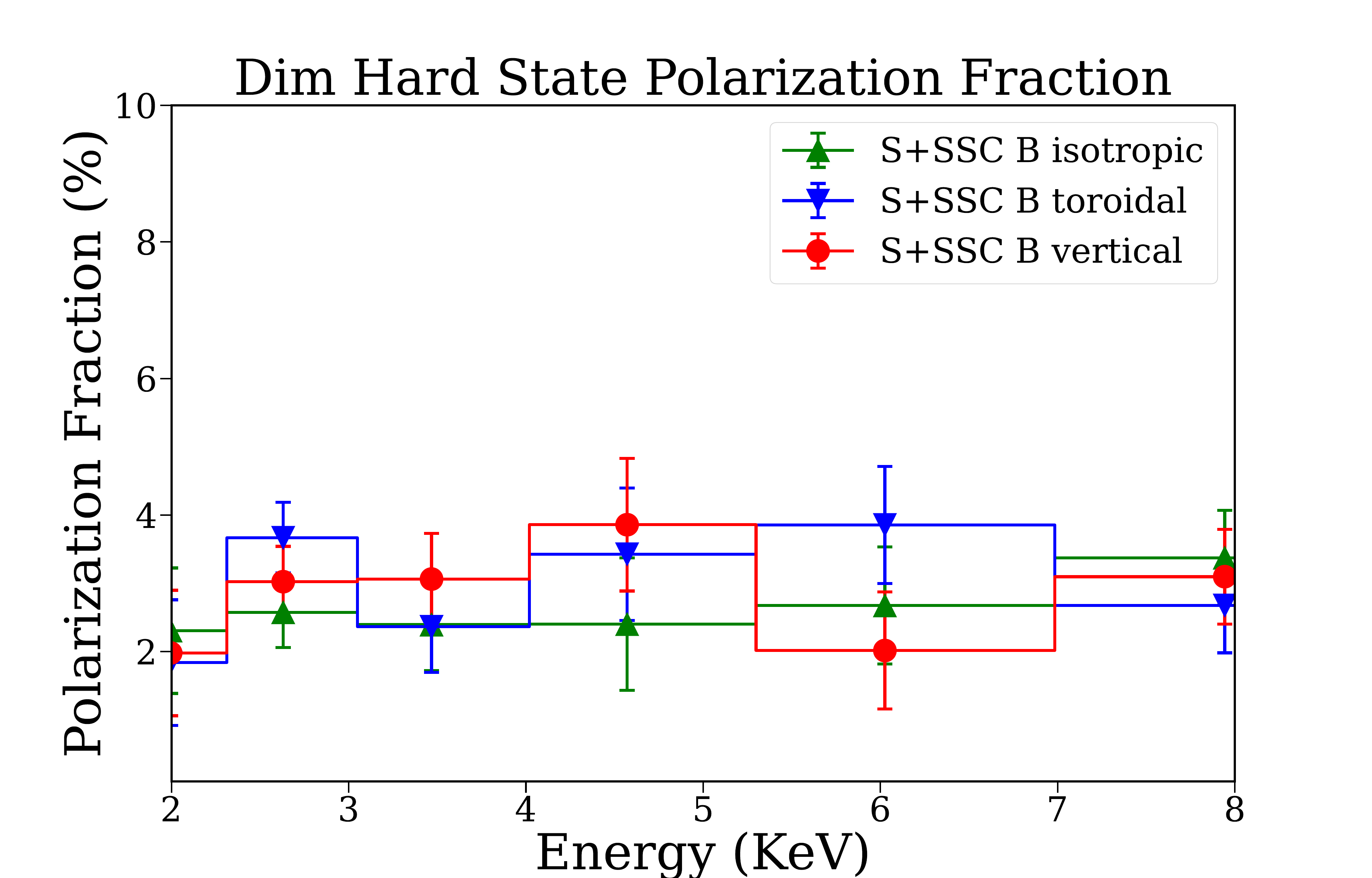}
\label{fig:Monk_bright_polfrac}

\includegraphics[width=0.75\columnwidth]{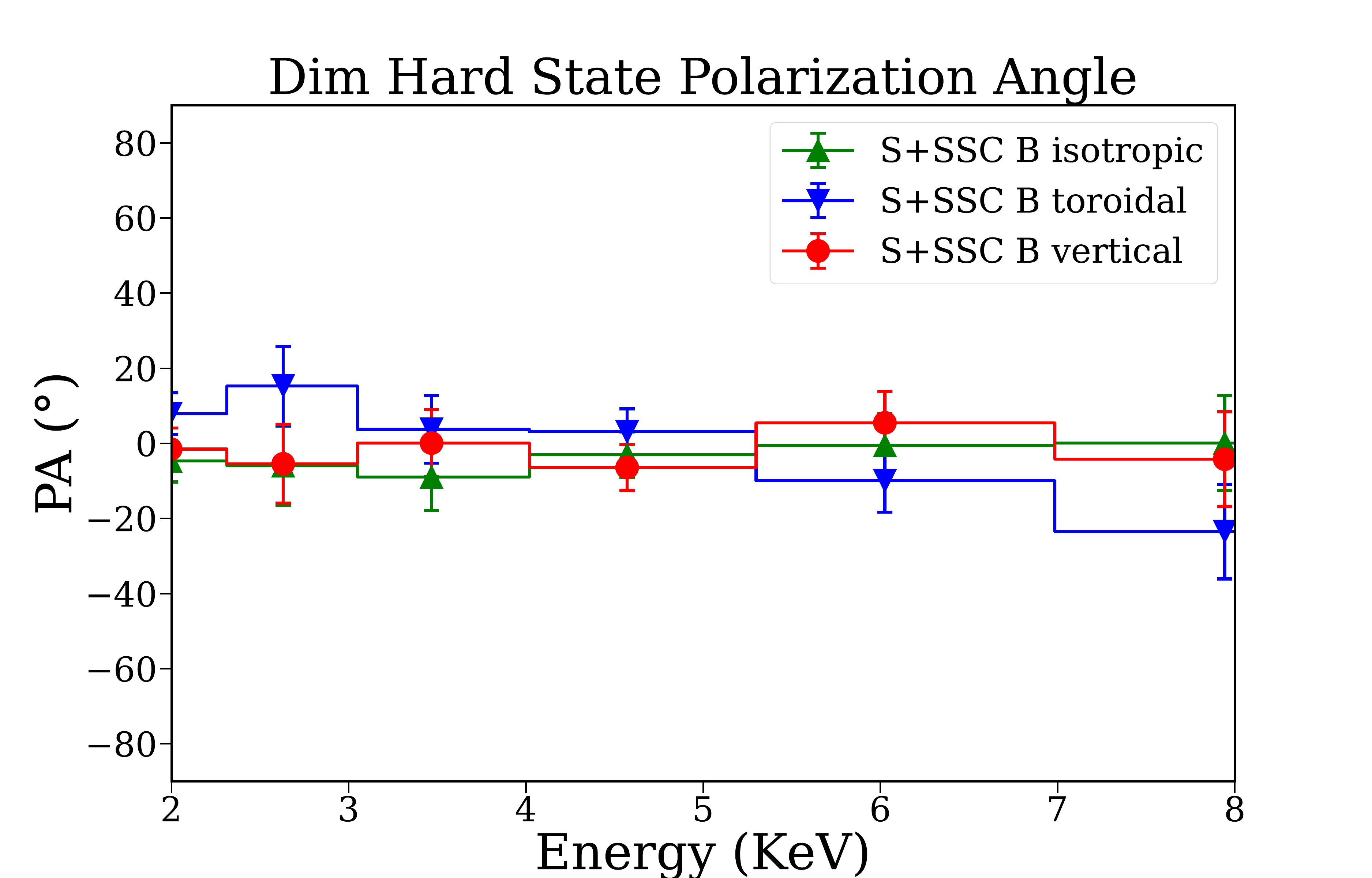}
\label{fig:Monk_bright_polang}

\caption{Dim hard state luminosity-weighted polarisation signature for different magnetic field configurations: vertical (red), toroidal (blue), and isotropic (green).}
\label{fig:pol_dim_b_config}
\end{figure}

\begin{figure}
\includegraphics[width=1.1\columnwidth]{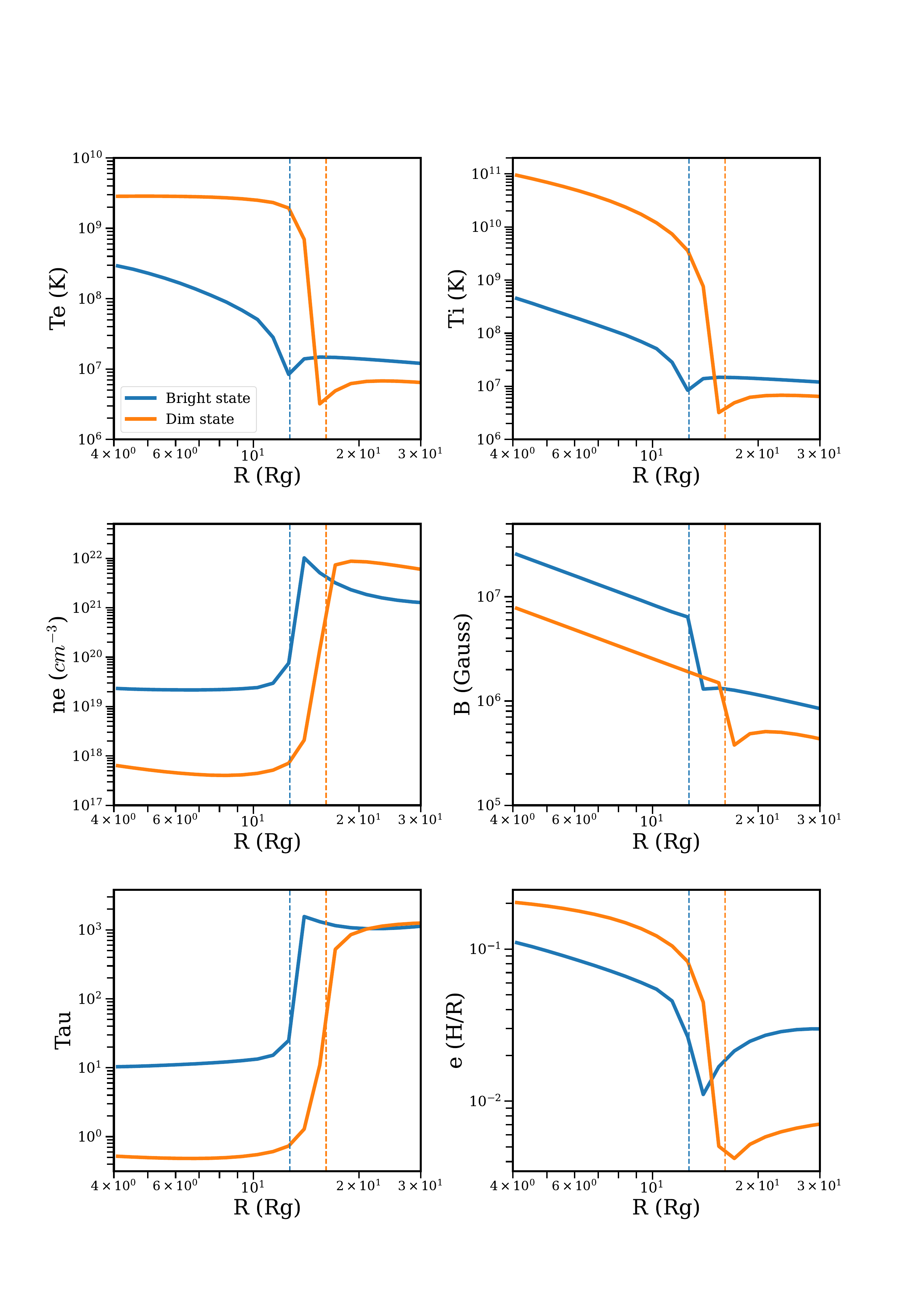}
 \caption{Radial profile of the disk parameters for the bright (in blue) and dim (in orange) hard-state JED--SAD configuration. The vertical line represents the transition radius ($R_J$) for each state. The parameters are: the electron temperature ($T_e$), the electron density ($n_e$), the vertical Thomson  optical depth ($\tau$), the disk aspect ratio ($H/R$), the magnetic field ($B$), and the ion temperature ($T_i$)} \label{fig:diskstruc}
\end{figure}

\FloatBarrier
\clearpage

\section{DYPLO versus MONK spectrum} \label{Appendix}

\begin{figure}[H]
\includegraphics[width=\columnwidth]{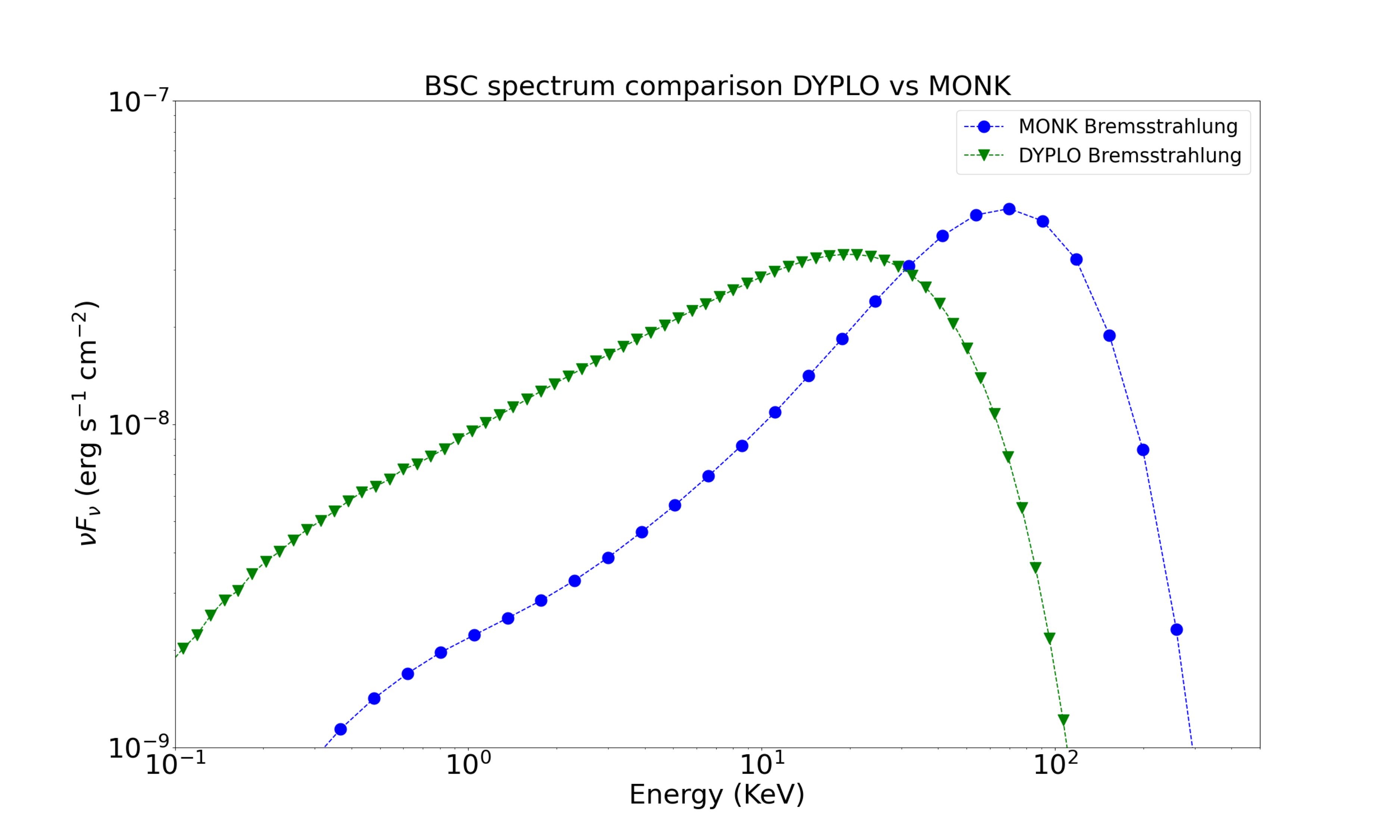}
 \caption{DYPLO BSC vs MONK BSC spectrum comparison for a set of JED--SAD parameters consistent with the bright hard state.  \label{fig:flux_comp_bright}}
\end{figure}

\begin{figure}[H]
\includegraphics[width=\columnwidth]{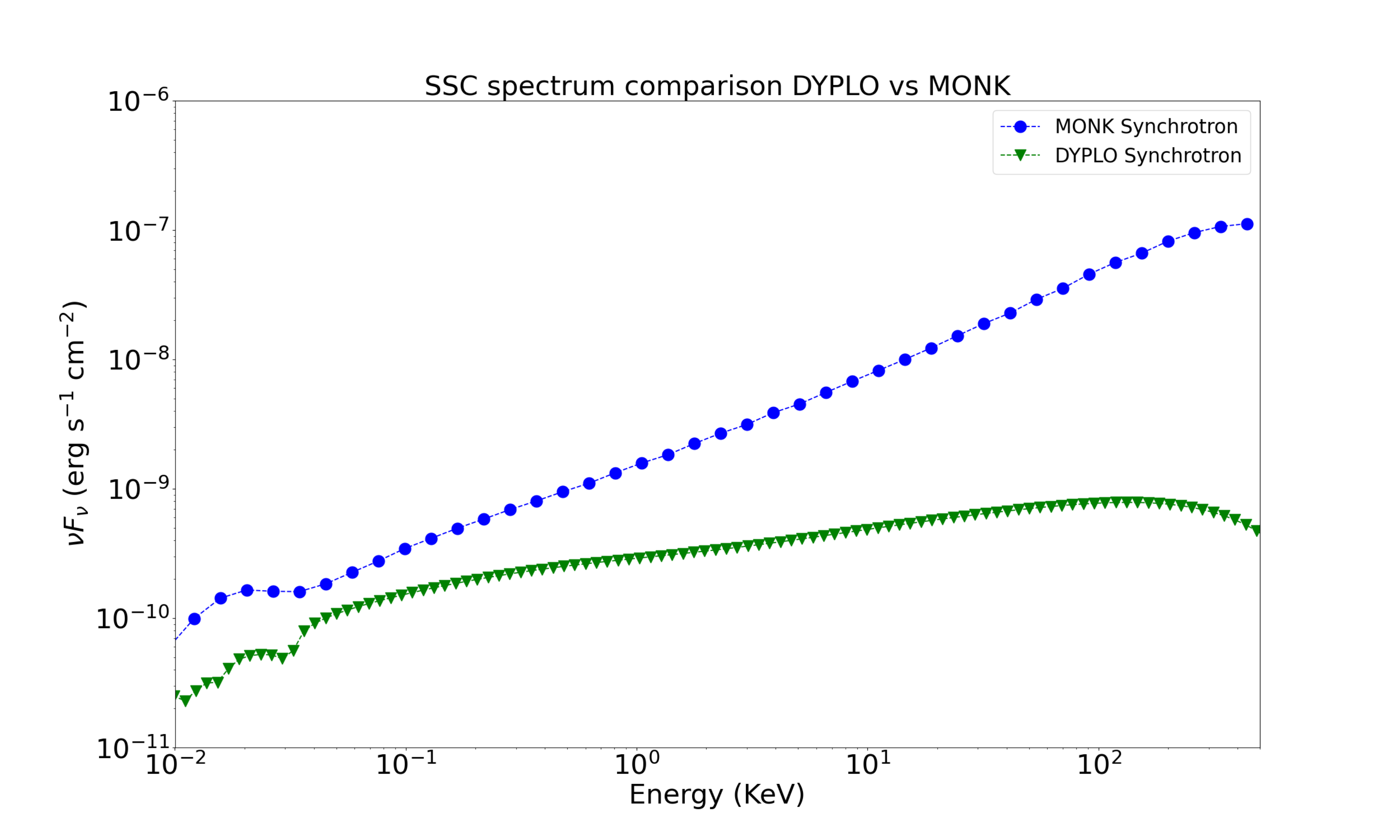}
 \caption{DYPLO SSC vs MONK SSC spectrum comparison for a set of JED--SAD parameters consistent with the dim hard state. \label{fig:flux_comp_dim}}
\end{figure}

During the comparison of the dominant spectral components derived from MONK and DYPLO, discrepancies were observed in the relative contributions of the emission processes (see Figs. \ref{fig:flux_comp_bright} and \ref{fig:flux_comp_dim}). To identify the origin of these differences, we performed a detailed, process-by-process analysis.

\subsection{Bright hard state, a bremsstrahlung-dominated regime}

\begin{figure}
\includegraphics[width=\columnwidth]{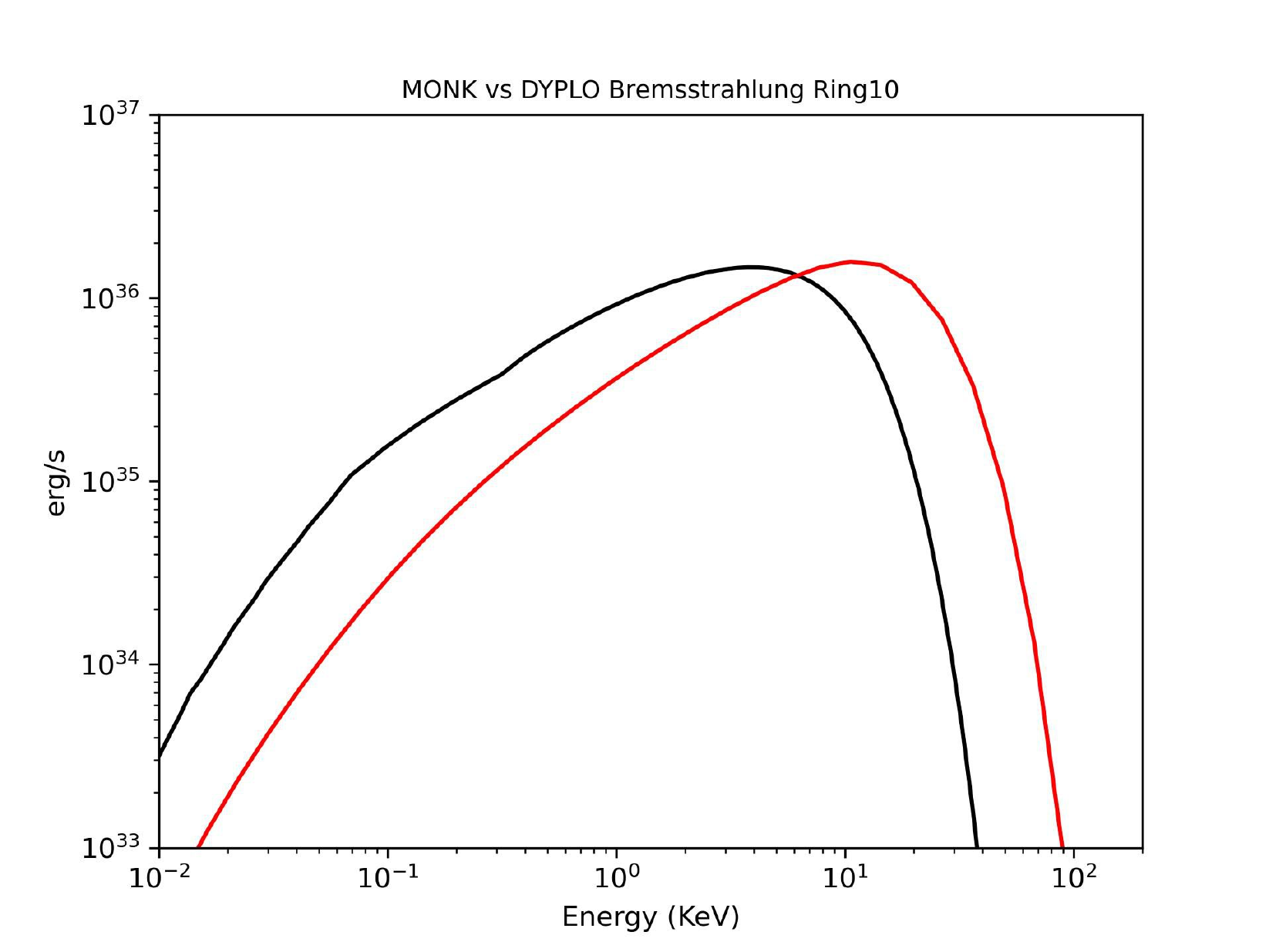}

\includegraphics[width=\columnwidth]{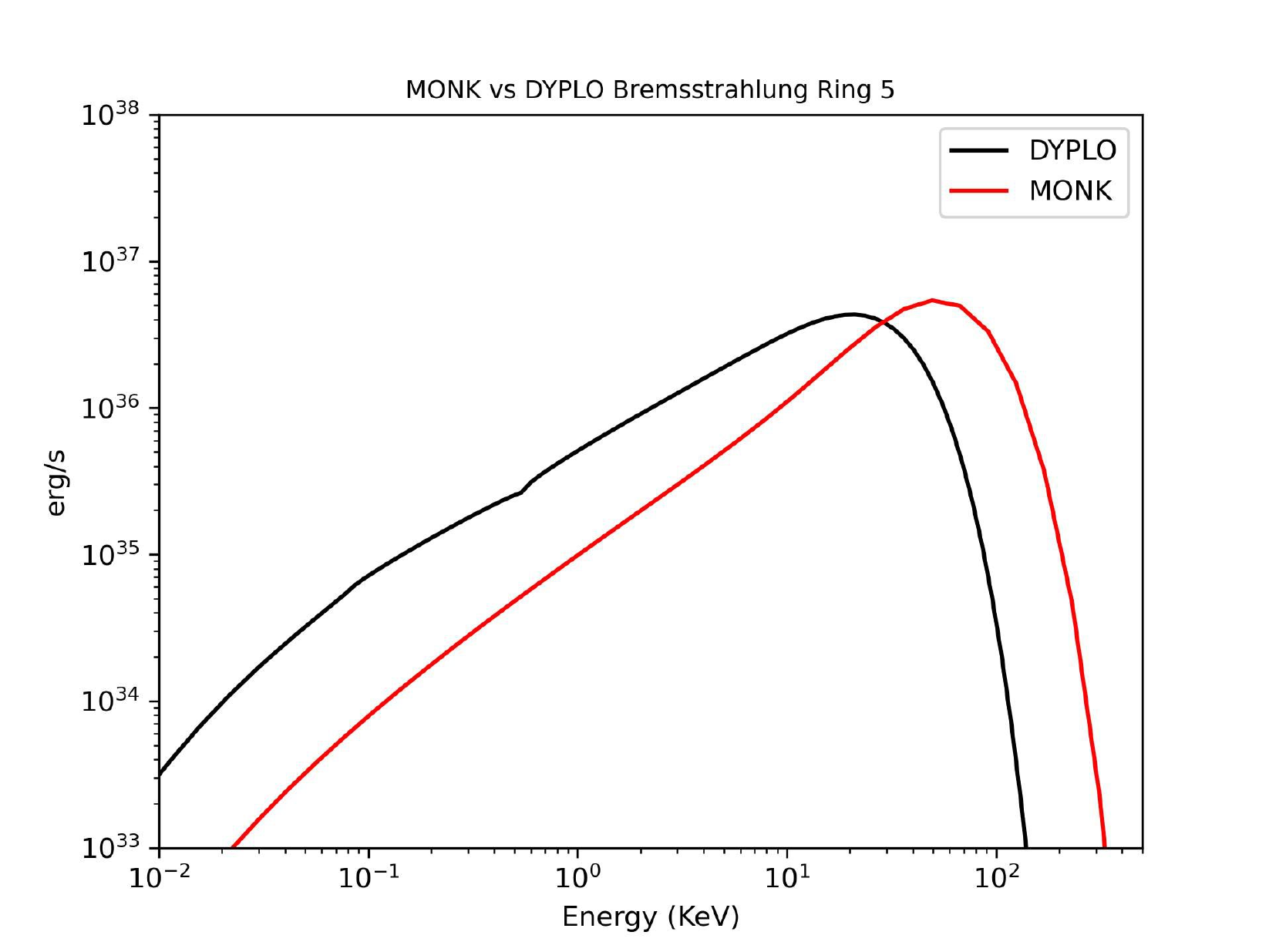}

\caption{MONK vs DYPLO BSC comparison for Rings 5 and 10. \label{fig:dypvsmonk_bright_init}}
 
\end{figure}

In the bright hard state, where the emission is dominated by bremsstrahlung, Fig. \ref{fig:dypvsmonk_bright_init} shows the bremsstrahlung contributions for several disk rings. Even within individual rings, significant differences are visible between the two codes. Possible causes included (i) the interpolation scheme used in DYPLO to read from the BELM tables, (ii) potential inconsistencies in the implemented formulae, or (iii) differences in the physical parameters considered during the computation.

After a systematic investigation, we determined that the source of discrepancy arises from the temperature definition used in each code. In MONK, the input temperature corresponds to the central temperature of the ring. In contrast, DYPLO, for optically thick solutions, uses an effective temperature derived from the bridging formula described in \cite{marcel_unified_2018}. This difference in temperature leads to distinct bremsstrahlung emissivities and therefore different spectral shapes.

\begin{figure}
\includegraphics[width=\columnwidth]{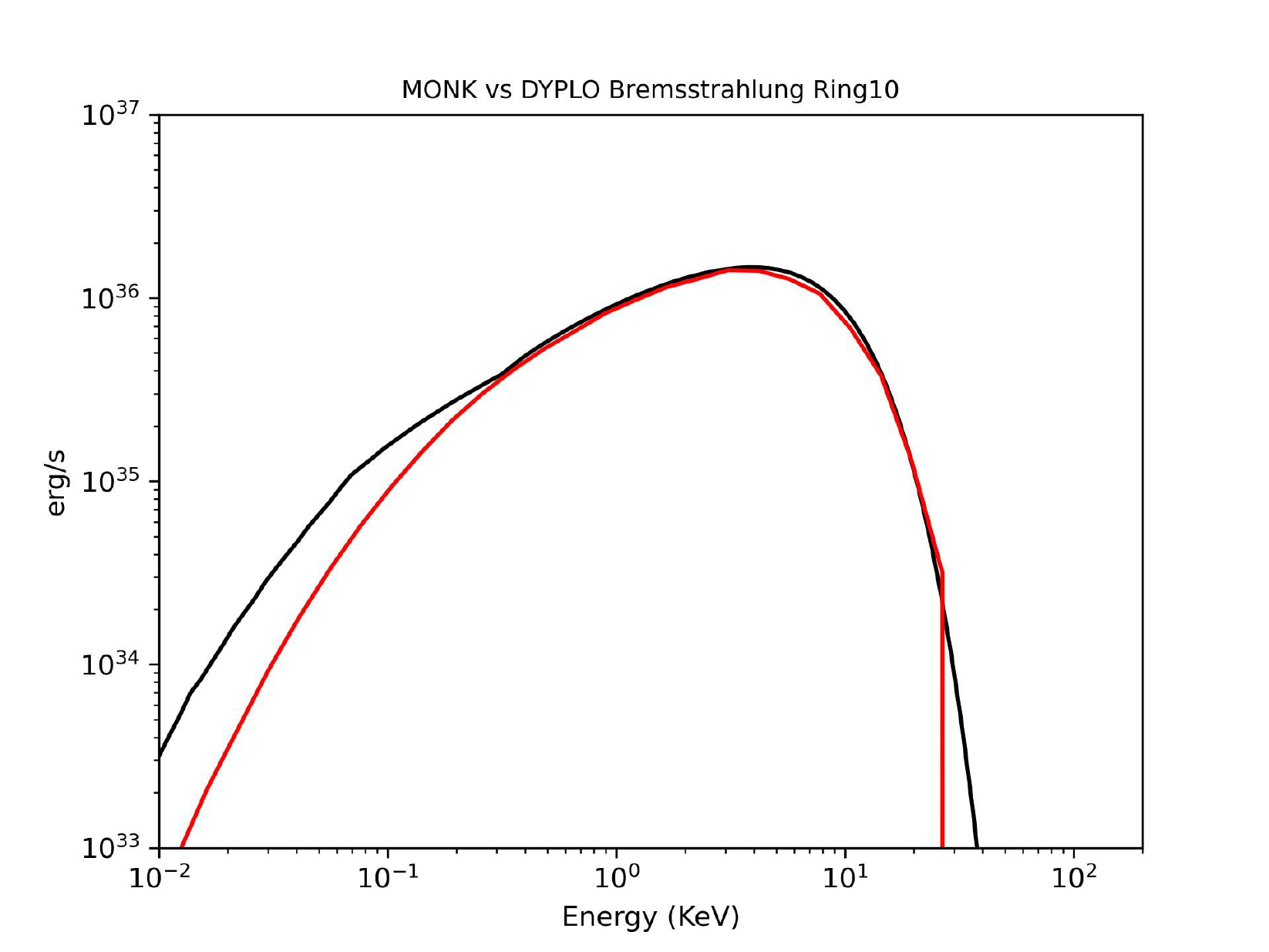}

\includegraphics[width=\columnwidth]{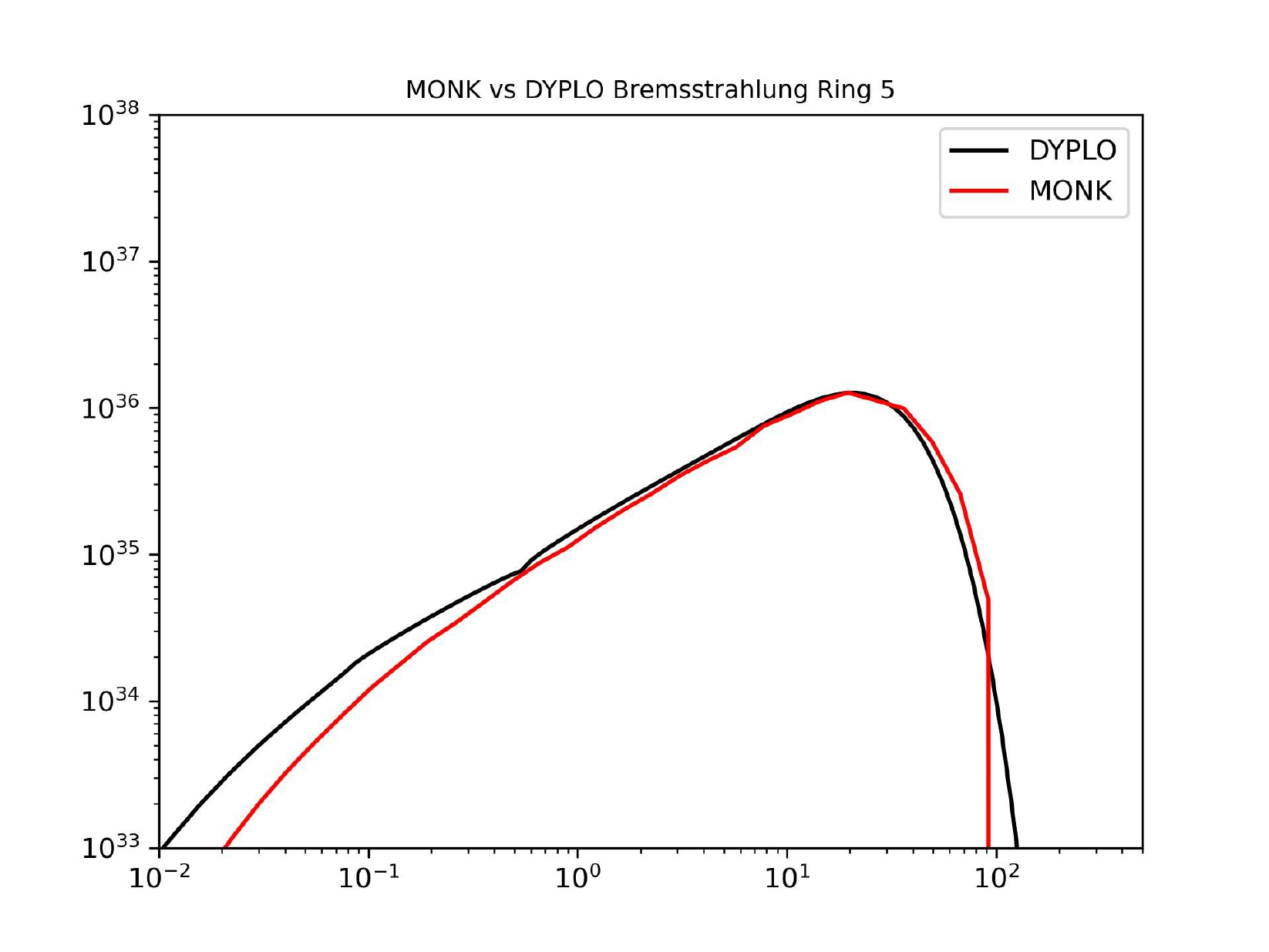}

\caption{MONK vs DYPLO BSC comparison with a consistent temperature for Rings 5 and 10. \label{fig:dypvsmonk_bremsstrahlung}}
 
\end{figure}

By implementing DYPLO's temperature in MONK the resulting spectra agreed closely (see Fig. \ref{fig:dypvsmonk_bremsstrahlung}), confirming that the temperature inconsistency was the primary cause of the observed differences. Note that some discrepancies remain at low energies. These are mainly due to known differences between the MONK and DYPLO calculations that will be detailed in a future work.

\subsection{Dim hard state, a synchrotron-dominated regime}

\begin{figure}
\includegraphics[width=\columnwidth]{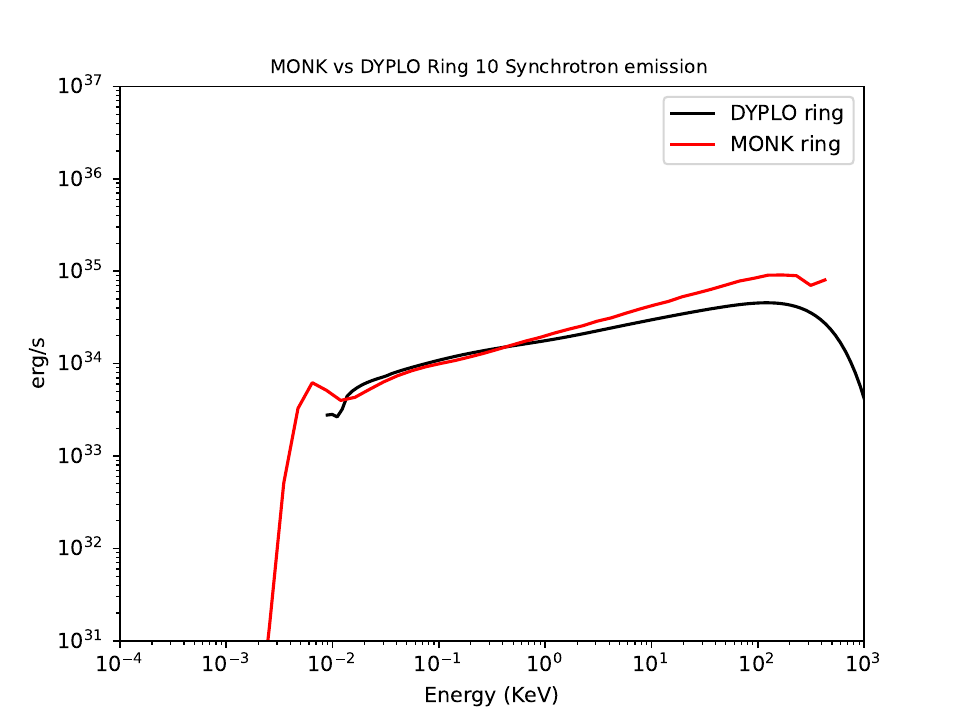}

\includegraphics[width=\columnwidth]{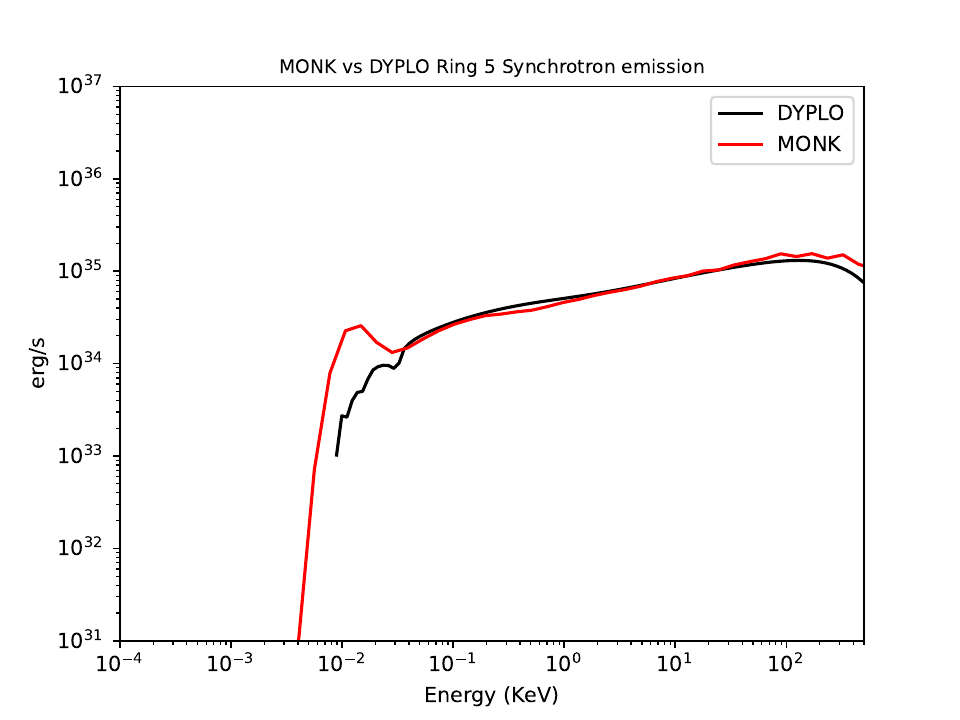}

\caption{MONK vs DYPLO SSC comparison for Rings 5 and 10. \label{fig:dypvsmonk_synchrotron}}
 
\end{figure}

In the dim hard state, where the emission is synchrotron-dominated, a pronounced difference was found in the spectral slope between MONK and DYPLO. To investigate this, we computed the synchrotron emission ring by ring (Fig. \ref{fig:dypvsmonk_synchrotron}). The slope differences do not appear on every ring spectrum, which we attributed to the interpolation of the BELM tables in DYPLO. However, these local discrepancies were insufficient to account for the overall difference in the total integrated spectrum.

\begin{figure}
\includegraphics[width=\columnwidth]{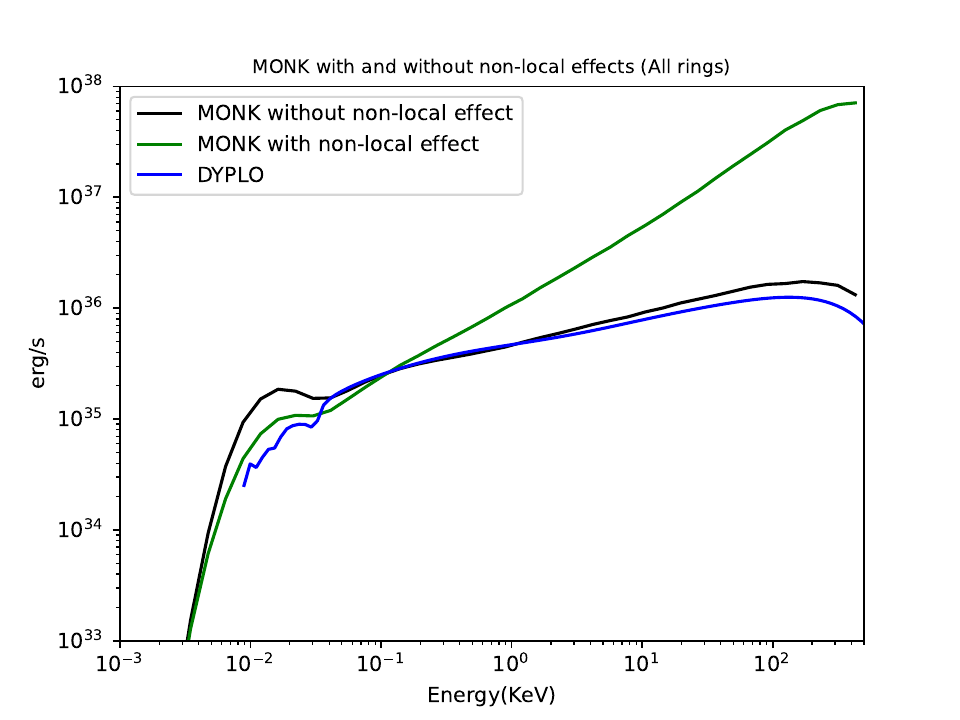}

\caption{MONK with non-local vs MONK without non-local vs DYPLO SSC dim hard state solutions. \label{fig:dypvsmonk_nonloc}}
 
\end{figure}

To investigate this discrepancy further, we computed the MONK spectrum of each individual ring separately, thereby isolating the emission originating from each ring. This procedure reproduces the emission calculation performed in DYPLO, where no radiative interactions between rings are taken into account. Figure~\ref{fig:dypvsmonk_nonloc} shows that this isolation process significantly alters the MONK spectrum. The resulting non-interacting MONK spectrum is in much better agreement with the DYPLO spectrum, indicating that non-local effects play a major role in this case. By non-local effects, we refer to photons produced at a radius R that propagate and interact at another radius R*
, a process that is not included in DYPLO, which solves the radiative equilibrium independently for each ring. As a Monte Carlo radiative transfer code, MONK naturally accounts for these non-local interactions.

This analysis demonstrates that the discrepancies between MONK and DYPLO in the synchrotron-dominated regime are primarily due to the non-local effect that is not taken into account in DYPLO. To achieve full consistency between the two codes, it is therefore necessary to implement MONK's cooling computation within DYPLO, ensuring a self-consistent treatment of radiative cooling and equilibrium across the two frameworks.

\end{appendix}

\end{document}